\documentclass[12pt]{article}
\usepackage{amsmath,amssymb,amsthm}
\usepackage{color}
\usepackage[nohead,margin=1.0in]{geometry}
\usepackage{booktabs}
\usepackage{algorithm,algorithmic}
\usepackage{verbatim}
\usepackage{url}
\usepackage{graphicx}
\usepackage{epstopdf}
\usepackage{mathtools}
\usepackage{tikz}
\usepackage[colorlinks=true, allcolors=blue]{hyperref}
\usetikzlibrary{spy,decorations.fractals,shapes.misc}
\mathtoolsset{showonlyrefs}

\graphicspath{{./pics/}}
\theoremstyle{plain}
\newtheorem{theorem}{Theorem}[section]
\newtheorem{remark}{Remark}[section]

\newtheorem{assumption}{Assumption}[section]

\numberwithin{equation}{section}
\providecommand{\argmax}{\operatorname*{argmax}} 				

\newcommand{\ceq}{\overset{\mathsf{c}}{=}}

\newcommand{\pll}{\varphi}
\newcommand{\sdir}{h}     
\newcommand{\ufrq}{\eta}  
\newcommand{\lgf}{\zeta}  
\newcommand{\fss}{s}      
\newcommand{\sss}{\tau}   
\newcommand{\gdir}{q}     

\title{Stochastic EM methods with Variance Reduction for Penalised PET Reconstructions}
\author{
 \v Zeljko Kereta\thanks{Department of Computer Sciences, University College London, Gower Street, London WC1E 6BT, UK. (\texttt{z.kereta@ucl.ac.uk}, \texttt{s.arridge@ucl.ac.uk}, \texttt{b.jin@ucl.ac.uk}). The work of ZK, SA, BJ is supported by UK EPSRC grant EP/T000864/1.}
 \and Robert Twyman\thanks{Institute of Nuclear Medicine, University College London, London, UK (\texttt{robert.twyman.18@ucl.ac.uk}, \texttt{k.thielemans@ucl.ac.uk}). The work of RT is supported by GE Healthcare, and of KT by EP/T026693/1.} \and Simon Arridge\protect\footnotemark[1] \and Kris Thielemans\protect\footnotemark[2] \and Bangti Jin\protect\footnotemark[1]
}

\begin{document}

\maketitle

\begin{abstract}
Expectation-maximization (EM) is a popular and well-established method for image reconstruction in positron emission tomography (PET)
but it often suffers from slow convergence. Ordered subset EM (OSEM) is an effective reconstruction algorithm that provides significant acceleration during initial iterations,
but it has been observed to enter a limit cycle. In this work, we investigate two classes of algorithms for accelerating
OSEM based on variance reduction for penalised PET reconstructions. The first is a stochastic variance reduced EM algorithm, termed as SVREM, an extension of
the classical EM to the stochastic context, by combining classical OSEM with insights from variance reduction techniques for gradient descent.
The second views OSEM as a preconditioned stochastic gradient ascent, and applies variance reduction
techniques, i.e., SAGA and SVRG, to estimate the update direction. We present several numerical experiments to illustrate the efficiency and accuracy of the approaches. The numerical results show that these approaches significantly outperform existing OSEM type methods for penalised PET reconstructions, and hold great potential.
\end{abstract}

\section{Introduction}\label{sec:intro}
Positron emission tomography (PET) is a nuclear imaging technique that allows for the measurement of biochemical changes in the body by observing the spatial distribution of a radioactive tracer.
Positron emitting radionuclides are attached to a biochemical compound to create a  radioactive tracer, e.g. fluorodeoxyglucose, that is used in natural metabolic processes by an organ or tissue of interest.
The radionuclides decay and the emitted positron travels a short distance before encountering and annihilating with an electron. This annihilation interaction results in a pair of 511 keV photons that travel anti-parallel. The emitted photons may be measured by a pair of detector elements
along a ring of crystalline detectors surrounding the subject. If two photons are detected within a short coincidence timing window, a PET scanner will record a coincidence event along the line-of-response between the two measuring detectors. The goal of PET image reconstruction problem is then to reconstruct an estimate of the emission distribution from the measured coincidence data. 
This inverse problem is ill-posed in the sense of Hadamard, i.e., the solution to the problem is not stable with respect to the perturbation in the data.

Iterative methods have been widely used in PET reconstruction \cite{QL06},
amongst which the expectation maximisation (EM) algorithm and its various variants, e.g., MLEM \cite{DLR77}, OSEM \cite{HL94}, RAMLA \cite{BdP96}, BSREM \cite{dPY01,AF03} and OS-SPS \cite{EF98},  are predominant. Shepp and Vardi \cite{SV82,SVK85}
reformulated the PET reconstruction problem into a maximum likelihood (ML) estimation of the tracer distribution,
and developed an iterative scheme via EM algorithm (MLEM), which enjoys several desirable features, e.g.,
a closed-form for iterate updates and nonnegativity preservation. The EM algorithm consists of two steps:
(i) the E-step computes the complete data sufficient statistic; and (ii) the M-step updates the estimate
by maximising the complete data log-likelihood. The algorithm converges monotonically (in decreasing the objective) but slowly. Moreover, full batch updates (i.e. using all measured data to compute the sufficient
statistic) can be costly for large data sizes. To mitigate the ill-posedness of the PET{} reconstruction problem,
a suitable penalty is employed, leading to a maximum a posteriori (MAP) problem
\cite{ItoJin:2015,HohageWerner:2016}. This requires adapting the standard EM algorithm, since 
a closed-form solution at the M-step is often no longer available \cite[p. R561]{QL06}. There are several approaches to address the challenge, such as the one-step-late algorithm \cite{G90} (applicable to differentiable
penalty terms but generally not convergent to the MAP solution), or more principled methods via
modified EM{} algorithms \cite{dP95} or separable parabolic surrogates for the penalty \cite{EF98}.

One established procedure to mitigate the aforementioned computational challenge with full batch data is the ordered subset EM{} (OSEM)
\cite{HL94}, which first divides the measured data into disjoint subsets and then applies the
EM{} algorithm to one subset at each iteration, either in a cyclic or a stochastic manner \cite{HL94}. This
greatly reduces the cost per update, and leads to significant
acceleration during initial iterations. However, the standard OSEM algorithm has been observed to often not converge but instead enters a limit cycle  \cite{B98, dPY01}. 
This has motivated intensive development of modified OSEM
algorithms that retain both the speed-up in early iterations but also exhibit convergence to the MAP solution 
(e.g., by suitably adjusting the step-size schedule) \cite{BdP96,dPY01,AF03}.

EM{} and OSEM{} can also be written in a gradient ascent-like form, where the search direction is a
preconditioned gradient of the objective \cite{Kaufman:1987}. This viewpoint enables 
designing a range of new methods that allow a general class of differentiable penalties.
This idea has recently been experimentally studied for PET{} with iteration-dependent and constant
preconditioners in \cite{TA+20} and \cite{TAT21}, respectively, both for standard stochastic ascent
approaches and for variance reduction methods. For nonsmooth convex penalties, e.g., total (generalised) 
variation \cite{BrediesHoller:2020}, gradient approaches are no longer directly applicable, and one 
may resort to a saddle point reformulation, and then update primal and dual variables accordingly \cite{CP11}. 
Chambolle et al. \cite{CE+18,EMS19} developed a stochastic variant of such an algorithm using only one random 
component of the dual variable at each update, and provide a convergence guarantee. Alternatively, one may employ proximal methods to handle non-smooth penalties, and many variance reduction algorithms have been extended to the proximal setting \cite{XZ14}.

In this work, we contribute to stochastic variance reduction algorithms for the MAP problem in PET, for
a popular class of penalty terms, by drawing on recent advances in stochastic optimisation and machine learning.
First, we develop a novel algorithm, termed as Stochastic Variance Reduced EM (SVREM), for the MAP reconstruction.
It is motivated by the online-EM \cite{OM09} (see also \cite{NealHinton:1999}) and its variance reduction variants 
\cite{CZ+18,KW+20}, originally developed for an un-penalised problem. We present an extension to the MAP problem 
by combining variance reduced EM for computing a variance reduced running average of the sufficient statistics with the surrogate approach for the penalty \cite{CAV04}. The resulting SVREM algorithm maintains the EM nature 
for PET reconstruction, e.g., nonnegativity preservation, and admits an explicit maximiser at each 
M-step. The overall algorithm is mathematically principled and numerically easy to implement. Second, we revisit variance reduction algorithms for stochastic gradient ascent, and their use in iterative PET reconstruction, which were recently experimentally studied \cite{TA+20,TAT21}. 
These algorithms do not belong to the EM family, but rather to the class of diagonally preconditioned gradient ascent 
algorithms. Due to the inclusion of the penalty, the non-negativity of the iterates is no longer ensured, which 
requires a projection step (i.e., the proximal map of the characteristic function on the set $\{f\geq0\}$). In Theorem \ref{thm:sagasvrg_convergence}, we show the almost sure convergence to a maximiser 
for a modified likelihood with a constant preconditioner. Third and last, we conduct extensive numerical experiments, 
which show that these algorithms enjoy steady convergence, significantly outperform classical OSEM type methods, and are very promising for PET reconstruction.

The rest of the paper is organised as follows. In Section \ref{sec:EM}, we describe the
mathematical formulation of the maximum likelihood problem in PET, and expectation
maximisation and its stochastic variants for ML estimation. Then in Section
\ref{sec:regularisation}, we discuss the EM algorithm for MAP reconstruction
using parabolic surrogates. In Section \ref{sec:GA}, we discuss a second class
of numerical algorithms, i.e., variance reduction algorithms based on gradient ascent for the MAP problem.
Last, in Section \ref{sec:numerics} we present numerical results that examine and illustrate
features of these algorithms. Throughout, the notation $\ceq$ denotes expressions that
are equal up to an additive constant that is independent of the function's argument.
The notation $\oslash$ and $\odot$ denote entrywise division and multiplication of
matrices or vectors. For any fixed $M\in\mathbb{N}$, we denote $[M]$ to be the set $[M]=\{1,\ldots,M\}$.
We use lowercase letters, e.g., $ g$ and $f$ for column vectors, and upper case letters for
matrices and operators. For a matrix $A$ we let $a_m$ denote its $m^{\text{th}}$ row.
The notation $P_{\geq0}(x)$ denotes the coordinate-wise projection of $x$ onto the
non-negative plane, i.e., $P_{\geq0}(x)=(\max(0,x_n))_{n=1}^N$. The notation $1$ is slightly 
abused for a constant vector of suitable size with all entries equal to one.


\section{Expectation maximisation and its stochastic variants}\label{sec:EM}

In this section, we describe the ML PET problem, and the EM algorithm and its stochastic variants for finding ML solutions.

\subsection{ML PET problem}
First we recall the standard mathematical formulation of the PET reconstruction problem.
Let $M$ denote the number of detector bins, and $ g_m$ the number of emissions detected in the
$m^{\text{th}}$ bin, so that the measured data are $ g = ( g_1,\ldots, g_M)^\top\in\mathbb{R}^M$.
It is customary to approximate the measurement means of the unknown tracer distribution in
the form of a linear problem
\begin{equation}\label{eqn:voxel_td}
\mathbb{E}[ g] = A  f +w,
\end{equation}
where $A\in\mathbb{R}^{M\times N}$ is the system matrix with non-negative elements, $ f\in\mathbb{R}^N$ is vector of
voxel values, and $w\geq 0$ represents the mean number of background events such as scatters, background radiation, and random coincidences, which will mostly not be explicitly written in the equations below.
Emission measurements in the $m^\text{th}$ bin are modelled by the following Poisson model:
$$ g_m\sim\text{Poisson}(\mathbb{E}[{ g}_m]).$$
Recall that a random variable $ g$ follows the Poisson distribution $\text{Poisson}(\lambda)$ with a parameter $\lambda>0$ if
\begin{equation*}
 \mathrm{Prob}( g=k) = \frac{\lambda^ke^{-\lambda}}{k!},\quad k=0,1,\ldots.
\end{equation*}
Assuming that detector bins record independent measurements, and conditioning on the tracer distribution $f$,
it follows that the probability distribution function $p(g|f)$ of the emission measurements $g$ is given by
\begin{equation}\label{eqn:pdf}
 p( g| f) = \prod_{m=1}^M \exp(-\mathbb{E}[ g_m]) \frac{{\mathbb{E}[ g_m]}^{ g_m}}{ g_m!}.
\end{equation}
The ML estimator $f_{\rm ml}$ of $ f$ is computed by maximising the likelihood $p(g|f)$ in
\eqref{eqn:pdf}, or equivalently its logarithm. Omitting terms independent of $ f$, this
yields the following objective
\begin{equation}\label{eqn:log_likelihood}
\mathcal{L}( f): = \log (p( g| f)) \ceq \sum_{m=1}^M (-a_m^\top  f +w_m+  g_m \log(a_m^\top  f+w_m)),
\end{equation}
The ML{} estimator $f_{\rm ml}$ is then defined as
\begin{equation}\label{eqn:ML} f_{\rm ml} \in \argmax_{ f\geq 0} \mathcal{L}( f).
\end{equation}
The functional $\mathcal{L}( f)$ is concave on the space of
all admissible tracer distributions ($ f\geq 0$), but a direct solution via Karush-Kuhn-Tucker
conditions is intractable, and instead iterative approaches are commonly used, which we discuss in more detail in Section \ref{ssec:EM} below.

Next we introduce the concept of ordered subsets.
Consider a partition $\mathcal{S}=\{S_1,\ldots,S_{N_s}\}$ of the set $[M]$, i.e. a collection of (sub)sets
such that $\emptyset\neq S_t\subset[M]$; $S_{t_1}\cap S_{t_2}=\emptyset$ for $t_1\neq t_2$;
and $\cup_{t=1}^{N_s} S_t = [M]$. For a vector $v$ and a matrix $A$, we denote by $v_t$ and
$A_t$ the subvector of length $|S_t|$  and an $|S_t|\times N$ submatrix whose
entries, respectively row indices, belong to $S_t$. Accordingly, given the partition $\mathcal{S}$, we can subdivide the
log likelihood $\mathcal{L}(f)$ into
\begin{equation}\label{eqn:subobjective}
\mathcal{L}( f) = \sum_{t=1}^{N_s} \mathcal{L}_t( f), \quad \mbox{with } \mathcal{L}_t( f) = \sum_{m\in S_t} (-a_m^\top  f+w_m +  g_m\log(a_m^\top f+w_m)).
\end{equation}
For many algorithms the partition $\mathcal{S}$ needs to be carefully constructed, in order to optimise the quality of the reconstructions \cite{HL94}.
Moreover, subsets should be balanced so that emission probabilities $\sum_{m\in S_t} a_{mn}$
are nearly independent of the subset index $t$ \cite{B98}, which is what we also adhere to. 
It is thus traditionally recommended
that iterating over subsets should follow an order such that projections corresponding
to next subset are as ``perpendicular'' as possible to previously used ones \cite{HM93}.

\subsection{ML expectation maximisation}\label{ssec:EM}
EM is the most well-known example of an iterative, functional substitution scheme for PET reconstruction.
It solves the ML problem \eqref{eqn:ML} by replacing the objective \eqref{eqn:log_likelihood} through a
complete data framework. We follow the complete data framework due to Shepp and Vardi \cite{SV82}.
Let $G\in \mathbb{R}^{M\times N}$ and $G_t\in\mathbb{R}^{|S_t|\times N}$ denote the full and subset complete data matrices,
respectively, with entries $ g_{mn}$ that denote the number of emissions detected in bin $m$ that originated from voxel site $n$.
Since
$$\mathbb{E}[ g_{mn}| f] =  a_{mn}f_n\quad \mbox{and}\quad \mathbb{E}[{ g}_m] = \sum_{n=1}^N \mathbb{E}[{ g}_{mn}],$$
the (subset) {complete data likelihood} $p(G_t|f)$ satisfies
\begin{equation}\label{eqn:complete_data_likelihood}
p(G_t| f) = \prod_{m\in S_t} \prod_{n=1}^N p( g_{mn}| f) = \prod_{m\in S_t}\prod_{n=1}^Ne^{-\mathbb{E}[ g_{mn}|f]}\frac{\mathbb{E}[ g_{mn}|f]^{ g_{mn}}}{ g_{mn}!}.
\end{equation}
Consider now the conditional expectation
\[
\mathbb{E}_{G| g, f}[\log p(G| f)]=\sum_{t=1}^{N_s}\mathbb{E}_{G_t| g_t, f}[\log p(G_t| f)],
\]
with
\begin{align}\label{eqn:EMsurrogate}
 \mathbb{E}_{G_t| g_t, f}[\log p(G_t| f)]
\ceq\sum_{n=1}^N e^t_{n}( f)f_n\log(f_n)  - f_n\sum_{m\in S_t} a_{mn}, \quad\text{and}\quad e^t_n ( f) = \sum_{m\in S_t} \frac{ a_{mn} g_m}{\sum_{l=1}^N  a_{ml}f_l}.
\end{align}
Each iteration of the{} OSEM{} algorithm consists of two steps:
\begin{description}
	\item[{E step.}] For a given subset index $t_k$, compute the expectation \begin{equation}\label{eqn:E-step}
	     \mathbb{E}_{G_{t_k}| g_{t_k}, f}[\log p(G_{t_k}| f)];
	\end{equation}
		\item[{M step.}] Maximise the expectation by \begin{equation}\label{eqn:Mstep} f^{(k+1)} = \argmax_{ f\geq 0} \mathbb{E}_{G_{t_k}| g_{t_k}, f}[\log p(G_{t_k}| f)] = \bigg( f_n^{(k)}e^{t_k}_n( f^{(k)})/\sum_{m\in S_{t_k}}  a_{mn}\bigg)_{n=1}^N.\end{equation}
\end{description}
When $N_s>1$, the above algorithm is referred to as OSEM{}.
The standard EM{} algorithm uses $N_s=1$ subset and obeys the update rule
\[
f^{(k+1)} =\bigg( f_n^{(k)} e_n( f^{(k)})/\sum_{m=1}^M  a_{mn}\bigg)_{n=1}^N, \quad \mbox{with }
e_n ( f) = \sum_{m=1}^M \frac{ a_{mn} g_m}{\sum_{l=1}^N  a_{ml}f_l}.
\]
From preceding equations it follows that EM{} and OSEM preserve nonnegativity of the updates.

The above framework falls under the umbrella of function substitution-type methods \cite{LRY00},
which at each step replace the original objective function with a surrogate.
Recall that a function $\widehat\Phi$ is said to be a surrogate of a concave objective $\Phi$ if it satisfies the following
properties:
\begin{equation*}
 \Phi( f) - \Phi( f^{(k)})\geq\widehat\Phi( f; f^{(k)})-\widehat\Phi( f^{(k)}; f^{(k)})
 \quad\mbox{and}\quad
\nabla \Phi( f)\vert_{ f= f^{(k)}} = \nabla \widehat\Phi( f; f^{(k)})\vert_{ f= f^{(k)}}.
\end{equation*}
These defining properties ensure that maximising the surrogate $\widehat\Phi$ monotonically increases the value
of the objective $\Phi$, thereby guaranteeing the convergence of the objective value. It can be shown that $\mathbb{E}_{G_t| g_t, f}[\log p(G_t| f)]$ is
a surrogate for $\mathcal{L}_t( f)$ \cite{LC84,SVK85,dP93,LRY00}.

\subsection{Stochastic Expectation Maximisation}\label{sec:sem}
The EM algorithm represents a powerful and versatile approach for inference and estimation involving distributions whose complete data likelihood 
belongs to the exponential family, i.e.,
\[
p(G| f)=c_{G}\exp(\lgf( f)^\top \fss(G) - \mathcal{U}( f)),
\]
\begin{equation}\label{eqn:expfam_terms}
\mathcal{U}( f) = \sum_{m=1}^M a_m^\top f,\,\,\,\lgf( f) = \big(\log(f_n)\big)_{n=1}^N,\,\,\,
\fss(G) = \bigg(\sum_{m=1}^M g_{mn}\bigg)_{n=1}^N.
\end{equation}
Then we can rewrite MLEM{} as
\begin{equation}\label{eqn:Statistic_EM}
f^{(k+1)}=\argmax_{ f\geq0} \lgf( f)^\top 
\fss( f^{(k)})-\mathcal{U}( f),
\end{equation}
where 
$$\fss(f^{(k)})=\mathbb{E}_{G\vert  g, f^{(k)}}[\log \fss(G)]$$ is the full sufficient statistic.
Physically, $s(G)\in\mathbb{R}^N$ represents the unknown emission quantities: the $n^{\text{th}}$ entry of the full sufficient statistics $\fss(G)$ is
$\sum_{m=1}^M g_{mn}$, the total number of emissions from the $n^{\text{th}}$ voxel.
Then we can interpret the E step in \eqref{eqn:E-step} as computing either the full expected statistic
or its subset variant.~A direct way to randomise OSEM is through a random sampling of the subsets.
This can be achieved by resampling the subset at each iteration, or choosing only the subset index
 at random (for a fixed partition $\mathcal{S}$). We employ the latter strategy since
in practice it shows superior performance \cite{TAT21}.

There have been several recent studies  \cite{ZW+17,OM09,CZ+18,KW+20} that randomise
the classical EM algorithm differently and show excellent performance on a range of problems, e.g.
Gaussian mixtures, natural language processing and hidden Markov models. One notable class
of these algorithms start from the expression \eqref{eqn:Statistic_EM} and instead of
computing the full expected statistics $\fss(f^{(k)})$ or the corresponding
subset statistics $\sss_{t_k}(f^{(k)})$ defined below, at each iteration use an exponentially running approximation
$\widehat{\fss}^{(k)}$. In the M-step we then compute
\begin{equation}\label{eqn:vrsemnoreg}
\widehat{f}^{(k)}=\argmax_{ f\geq 0} \lgf( f)^\top \widehat{\fss}^{(k)}-\mathcal{U}(f).
\end{equation}
To compute the estimate $\widehat{\fss}^{(k)}$, which approximates $\fss(f^{(k)})$, the
common practice is to view the full statistics as an average of $N_s$ subset statistics
\begin{align*}
\fss(f) = \frac{1}{N_s}\sum_{t=1}^{N_s} \sss_t(f),
\end{align*}
with the subset statistics $\sss_t$ given by
\begin{equation}\label{eqn:subsetstat}
\sss_t( f) = N_s  f\odot(\nabla\mathcal{L}_t( f)+A_t^\top{1}),\quad t\in [N_s].
\end{equation}
Both $\fss(G)$ and $\sss_t$ are both of the size of $f$. Then at each iteration, we (randomly) select an index $n$ to update the estimate $\widehat{\fss}^{(k)}$ (to $\fss(f^{(k)})$.

This idea was initially proposed to derive an online EM algorithm \cite{NealHinton:1999,OM09} for handling streaming data,
in order to approximate the conditional statistics by exponentially moving averages as 
the data streams in. The resulting updates are akin to subset gradient updates. Specifically,
for an initial guess $\widehat{\fss}^{(0)}$ and a subobjective index $t_k$, it can be written as
\begin{align}\label{eqn:tempsem}
\widehat{\fss}^{(k+1)} = (1-\alpha_k)\widehat{\fss}^{(k)} +\alpha_k \sss_{t_k}({\widehat f}_{\mathrm{sem}}^{(k)}),
\end{align}
where $\{\alpha_k\}$ is a decaying stepsize schedule, and the index $t_k$ is drawn
uniformly at random. This algorithm is termed as Stochastic Expectation Maximisation (SEM) below. Similar to classical stochastic gradient
descent algorithms \cite{BottouCurtisNocedal:2018}, the convergence of the resulting sequence of iterates  ${\widehat f}_{\mathrm{sem}}^{(k)}$
is highly dependent of the variance of the estimated statistics $\sss_{t_k}(f^{(k)})$ (compared with $\fss(f^{(k)})$ of the standard MLEM
algorithm). Thus, the convergence guarantee requires a decaying stepsize schedule.

To reduce the variance of the gradient estimate (for stochastic optimisation) and to allow a constant stepsize, in recent years, several variance reduction techniques
have been developed in the machine learning community, e.g., SAG \cite{SRB17}, SAGA 
\cite{DBLJ14}, SVRG \cite{JZ13}, and SARAH \cite{NL+17}; see \cite{GowerSchmidtBach:2020}
for an up-to-date overview. These techniques reduce the variance of the gradient by including in the 
search direction an average of the full gradient, which is updated either according to a 
predefined update schedule, or per-iteration. For Online-EM, they can be used to reduce the 
variations of the sufficient statistics estimate. A variant of online EM,
inspired by SVRG, was developed in \cite{CZ+18}, which uses an anchor point $f^{\text{anc}}$ and a full but 
infrequently updated estimate of $\fss(f)$ at $f^{\text{anc}}$:
\begin{align}\label{eqn:tempsvrem}
\widehat{\fss}^{(k+1)} &= (1-\alpha)\widehat{\fss}^{k} +\alpha\Big( \sss_{t_k}({\widehat f}_{\mathrm{svrem}}^{(k)}) - \sss_{t_k}({\widehat f}^{\text{anc}}) + s^{\text{anc}}\Big)\\
\text{If } k\,\, {\rm mod}\,\, \ufrq N_s &= 0, \text{ set }  f^{\text{anc}}=  f_{\mathrm{svrem}}^{(k)} \text { and update } s^{\text{anc}}=\fss(f^{\text{anc}}),
\end{align}
The full expectation $\fss^{\text{anc}}$ and the anchor estimate $f^{\text{anc}}$, are updated 
once every $\ufrq\in\mathbb{N}$ epochs, where one epoch refers to every $N_s$ iterations. It is worth noting 
that for both online-EM and its variance reduced variants, the resulting updates still follow the EM 
paradigm, where the M-step uses the computed running estimate of the sufficient statistic.

\begin{remark}
Naturally all other variance reduction techniques can be applied to improve SEM.
For example an algorithm based on SAGA reads \cite{KW+20}
\begin{align*}
\begin{split}
\widehat{\fss}^{(k+1)} &= (1-\alpha)\widehat{\fss}^{k} +\alpha\bigg(\sss_{t_k}( f^{(k)})- \fss^{(k)}_{t_k}+\frac{1}{N_s}\sum_{t=1}^{N_s}\fss^{(k)}_{t}\bigg)\\
\text{draw } {\tilde t_k}\in&[N_s] \text{ and set }\fss^{(k+1)}_{\tilde t_k}=\sss_{\tilde t_k}( f^{(k)}), \text{ and for } t\neq \tilde t_k \text{ keep } \fss^{(k+1)}_{t}=\fss^{(k)}_{t};\\
\end{split}
\end{align*}
We shall not examine these variants, and focus only on SVREM.
\end{remark}

The methodology described above naturally extends to the ordered subset framework for the PET problem
\eqref{eqn:log_likelihood}. Indeed, with the background events included, we can obtain
\begin{align}\label{eqn:tau_t}
    \sss_t( f)=\mathbb{E}_{G_{t}\vert  g_{t}, f^{(k)}}[\log \fss(G_t)]=N_s\sum_{m\in S_t}\bigg(\frac{ g_m a_{mn}f_n}{\sum_{\ell=1}^N a_{ml}f_l+w_m}\bigg)_{n=1}^N = N_s \, f\odot A_t^\top( g_t\oslash(A_t  f+w_t)).
\end{align}
For any estimator $\hat{\fss}^{(k)}$ of the full statistic $\fss(f^{(k)})$, for the PET ML problem \eqref{eqn:ML}, the maximisation
step in equation \eqref{eqn:vrsemnoreg} still admits a unique solution and can be computed as \eqref{eqn:Mstep}.

Now we briefly comment on the convergence of SVREM \eqref{eqn:tempsvrem}. The convergence result in
\cite{CZ+18} requires the subset statistics $\sss_t( f)$ to be Lipschitz continuous,
which holds only for nonzero backgrounds $w_i$. This condition arises also for
standard MLEM{} \cite{AF03}. There are two common remedies, a practical and a theoretical one.
The former is to set the pixel value to $0$ whenever the denominator in \eqref{eqn:EMsurrogate}
or \eqref{eqn:Mstep} is equal to zero. The latter is to modify the likelihood term, using a
quadratic approximation near the origin \cite{AF03}. Specifically, let $\pll_m(\ell)= g_m \log(\ell)-\ell,$
and we define
\begin{equation}\label{eqn:modified_LL}
\hat \pll_m(\ell)=\begin{cases}
\frac{\pll_m''(\varepsilon)}{2}(\ell-\varepsilon)^2+\pll'_m(\varepsilon)(\ell-\varepsilon)+\pll_m(\varepsilon), &\text{if } \ell\leq \varepsilon \text{ and } w_m=0\\
\pll_m(\ell),&\text{otherwise}
\end{cases},
\end{equation}
with the constant $\varepsilon>0$. If $\varepsilon$ is chosen to be sufficiently small, then the solution set does not change \cite{AF03}.

\section{Stochastic variance reduced EM (SVREM) for penalised PET reconstruction}\label{sec:regularisation}
To address the inherent ill-posed nature of the ML problem \eqref{eqn:ML}, one popular approach
is variational regularisation, which introduces a convex penalty $\mathcal{R}(f)$  \cite{ItoJin:2015}. This can often be interpreted as
a maximum a posteriori (MAP) estimation and the corresponding estimator $f_{\rm map}$ is given by
\begin{equation}\label{eqn:PET-MAP}
  f_{\rm map}=\argmax_{ f\geq 0}\{\Phi( f):=\mathcal{L}( f) - \beta \mathcal{R}( f)\}.
\end{equation}
A common type of penalties used in PET reconstruction take the form
\[ \mathcal{R}( f) = \frac{1}{2}\sum_{n=1}^{N}\sum_{j\in\mathcal{N}_n}w_{nj}\,\rho\big(f_n-f_j\big),\]
where $w_{nj}\geq0$ are weights, $\mathcal{N}_n$ is the $n^\text{th}$ voxels neighbourhood, 
and $\rho$ is a {potential} function. The penalty $\mathcal{R}(f)$
is used to promote the desired image structure, which are often meant to be locally smooth but still preserve edge phenomena.
Thus $\rho$ should smooth within a given tissue or organ, while retaining sharp boundaries between different tissues. Most smooth potentials are thus
monotonic, non-decreasing functions of the intensity difference $|f_n-f_j|$ that are roughly quadratic near
the origin and linear away from the origin, and satisfy the following assumption.
\begin{assumption}\label{asn:potentials}
The potential $\rho(f)$ is symmetric, everywhere continuously differentiable, with $\rho'(f)$ non-decreasing (so that $\rho(f)$ is convex).
The curvature function $\gamma_\rho(f) = \frac{\rho'(f)}{f}$ is assumed to be non-increasing for $f\geq0$, and such that $\lim_{f\searrow0}\gamma_\rho(f)$ is finite and non-zero.
\end{assumption}
A list of commonly used penalty terms satisfying these properties is given in Table \ref{tab:regularisers}. Note that this does not cover the relative difference penalty \cite{NB+02}.
For the \textsf{Huber}, \textsf{log cosh} and \textsf{hyperbola} penalties, the parameter $\delta>0$
controls the transition between the quadratic (smooth) and linear (edge-preserving) regimes of the given penalty term.

\begin{table}[!htbp]
\begin{center}
\caption{Commonly used penalty terms}\label{tab:regularisers}
\begin{tabular}{@{}ccccc@{}}
\toprule
      & $\rho(f)$ & $\rho'(f)$ & $\gamma_\rho(f)$ & {comments} \\ \toprule
      {quadratic}& $\frac{f^2}{2}$&$f$&1&{\footnotesize{not edge preserving}}\\
      {Huber}&$\begin{cases}\frac{f^2}{2},&|f|\leq\delta\\ \delta |f|-\frac{\delta^2}{2}, &|f|\geq\delta\end{cases}$  &$\begin{cases}f,&|f|\leq\delta\\{\delta}, &|f|\geq\delta\end{cases}$&$\begin{cases}1,&|f|\leq\delta\\\frac{\delta}{|f|}, &|f|\geq\delta\end{cases}$&{\footnotesize{not strictly convex nor }$\mathcal{C}^2$}\\
      {log cosh}&$\delta^2\log\cosh(f/\delta)$&$\delta\,{\tanh({f}/\delta)}$&$\delta\frac{\tanh({f}/\delta)}{{f}}$&\\
      {hyperbola}&$\delta\big(\sqrt{1+(f/\delta)^2}-1\big)$&$\frac{f}{\sqrt{1+(f/\delta)^2}}$&$\frac{1}{\sqrt{1+(f/\delta)^2}}$&{\footnotesize{approximates TV}}\\
\bottomrule
\end{tabular}
\end{center}
\end{table}

Note that the general principle for solving the corresponding MAP{} problem
\eqref{eqn:PET-MAP} by EM{} does not change. That is,
for MLEM, OSEM, SVREM, and SEM, instead of \eqref{eqn:vrsemnoreg}, we compute
\begin{equation}\label{eqn:vrsemreg}
\widehat{ f}^{(k)}=\argmax_{ f\geq\boldsymbol{0}} \lgf( f)^\top \widehat{\fss}^{(k)}-\mathcal{U}(f)-\beta\mathcal{R}(f).
\end{equation}
If the penalty $\mathcal{R}( f)$ is separable
(i.e., no coupling between the entries, which for example is the case for the quadratic prior), the objective function in \eqref{eqn:vrsemreg} is
separable and the M-step has a closed-form solution. However, this is not the case for most penalties
of interest in PET reconstruction and computing the maximiser requires solving a coupled system of equations.
Thus, \eqref{eqn:vrsemreg} is often maximised iteratively \cite{QL06}.

To explicitly solve the M-step, we employ a separable surrogate of the penalty $\mathcal{R}(f)$.
Surrogates have been widely used in penalised PET reconstruction  \cite{dP95,dPY01,FH95,EF98}.
The idea is to construct a surrogate for the potential $\rho$, either to facilitate the computation
of the prior or to improve conditioning (and convergence). We employ
the parabolic surrogate defined in \cite{CAV04}. Namely, consider the surrogate for the potential $\rho$ given by
\begin{align*}
\widehat\rho^{(k)}(f_n;f_j)&=\rho(f_n^{(k)}-f_j^{(k)})+\rho'(f_n^{(k)}-f_j^{(k)})\big(f_n-f_j-(f_n^{(k)}-f_j^{(k)})\big)\\
   & \quad + \gamma_\rho(f_n^{(k)}-f_j^{(k)})\big((f_n-f_n^{(k)})^2+(f_j-f_j^{(k)})^2\big)\\
&\ceq\gamma_\rho(f_n^{(k)}-f_j^{(k)}) \big( \big(f_n-\tfrac{f_n^{(k)}+f_j^{(k)}}{2}\big)^2+\big(f_j-\tfrac{f_n^{(k)}+f_j^{(k)}}{2}\big)^2\big),
\end{align*}
and define the surrogate penalty by
\[ \widehat{\mathcal{R}}( f; f^{(k)}) = \frac{1}{2}\sum_{n=1}^{N}\sum_{j\in\mathcal{N}_n}w_{nj}\,\widehat\rho^{(k)}(f_n; f_j).\]
Then for $n\neq j$ the $n^{\text{th}}$ and the $j^{\text{th}}$ entry are decoupled since the partial derivative $\frac{\partial \widehat\rho^{(k)}(f_n; f_j)}{\partial f_n}$ are given by
\[ \frac{\partial \widehat\rho^{(k)}(f_n; f_j)}{\partial f_n} = \gamma_\rho(f^{(k)}_n-f^{(k)}_j)(2f_n-f_n^{(k)}-f_j^{(k)}).\]
The $n^\text{th}$ partial derivative for the  surrogate objective $\lgf( f)^\top \widehat{\fss}^{(k)}-\mathcal{U}( f)-\beta\widehat{\mathcal{R}}( f; f^{(k)})$,  is given by
\begin{equation}\label{eqn:svrsurder} \frac{1}{f_n} \widehat{\fss}_n^{(k)}-2\beta f_n \sum_{j\in\mathcal{N}_n} d_{nj} +\bigg(\beta f_n^{(k)}\sum_{j\in\mathcal{N}_n} d_{nj} +\beta\sum_{j\in\mathcal{N}_n} d_{nj} f_j^{(k)}-\sum_{m=1}^M a_{mn}\bigg),
\end{equation}
where $\widehat s^{(k)}$ is an estimator of the expected statistic, and $d_{nj}:=w_{nj}\gamma_\rho(f^{(k)}_n-f^{(k)}_j)$.
Equating \eqref{eqn:svrsurder} with zero gives a scalar equation of the form $$\frac{a}{f}-2bf+c=0,$$
with
\begin{align*}
a&=\widehat{\fss}_n^{(k)}, \quad  b=\beta\sum_{j\in\mathcal{N}_n} d_{nj},\quad
c=\beta f_n^{(k)}\sum_{j\in\mathcal{N}_n} d_{nj} +\beta\sum_{j\in\mathcal{N}_n} d_{nj} f_j^{(k)}-\sum_{m=1}^M a_{mn}.
\end{align*}
Thus, we arrive at a quadratic equation, with a unique nonnegative solution which can be easily evaluated at each iteration.
Provided that $b>0$ it is given as
\[f =\frac{c+\sqrt{c^2+8ab}}{4b},\]
where we note that the discriminant is non-negative due to Assumption \ref{asn:potentials},
and when $b=0$, i.e. when there is no prior, equation \eqref{eqn:svrsurder} is linear, and the solution is given as $f = -c/a$.

\section{Stochastic EM algorithm based on gradient ascent}\label{sec:GA}
In this section, we describe a second class of algorithms for problem \eqref{eqn:PET-MAP}.
It is inspired by the following additive formulation of the EM update \eqref{eqn:Mstep}
\cite{Kaufman:1987}:
\begin{equation}\label{eqn:osem_additive}
 f_{\rm osem}^{(k+1)} =  f_{\rm osem}^{(k)} + ({ f_{\rm osem}^{(k)}}\oslash {A_{t_k}^\top {1}})\odot\nabla \mathcal{L}_{t_k}( f_{\rm osem}^{(k)}),
 \quad \mbox{with }\nabla \mathcal{L}_{t}( f) = A_t^\top( g_t\oslash(A_t  f)-{1}).
\end{equation}
This can be interpreted as a preconditioned form of gradient ascent. Then it is natural
to replace the gradient of the likelihood $\mathcal{L}(f)$ with that of
$\Phi(f)$. This strategy is directly amenable to stochastic iterative methods, which
are very appealing due to their low cost per iteration, flexibility with the penalty (i.e., there is no need for surrogates but only the gradient of the penalty) and the convergence acceleration during
early iterations. Specifically, stochastic gradient ascent (SGA) like methods can be written as
\begin{equation*}
   f^{(k+1)} =  f^{(k)} + \alpha_k \sdir_k( f^{(k)},\xi_k),
\end{equation*}
where the random (index) variable $\xi_k$ may depend on $ f^{(k)}$, 
and $\sdir_k( f^{(k)}, \xi_k)$ is
the search direction (i.e., a preconditioned version of the gradient of $\mathcal{L}(f^{(k)})$. Then OSEM can be written in the additive formulation
\eqref{eqn:osem_additive} with the search direction $ \sdir_k( f^{(k)},t_k)$ given by
 \begin{equation}
h_k( f^{(k)},t_k) =({ f^{(k)}}\oslash {A_{t_k}^\top {1}})\odot\nabla \mathcal{L}_{t_k}( f^{(k)}).
\end{equation}
This can interpreted as a diagonally preconditioned gradient ascent with respect to the
subobjective $\mathcal{L}_{t_k}$. These discussions naturally motivate the following algorithmic developments for the PET MAP problem \eqref{eqn:PET-MAP}. For a given subset index $t$, we denote
\[
\Phi_t( f):= \mathcal{L}_t( f) - \frac{\beta}{N_s} \mathcal{R}( f).
\]
Then the additive formulation can be extended to MAP estimation, which leads to SGA updates
\begin{equation}\label{eqn:sga}
 f_{\rm sga}^{(k+1)} =  f_{\rm sga}^{(k)} + \alpha_k d_t({ f_{\rm sga}^{(k)}})\odot \Big(\nabla\mathcal{L}_{t_k}( f_{\rm sga}^{(k)})-\frac{\beta}{N_s}\nabla\mathcal{R}( f_{\rm sga}^{(k)})\Big),
\end{equation}
where the index $t_k$ is selected uniformly at random, $\alpha_k=1$,
and the preconditioner $d_t(f)$ is given by
\begin{equation}\label{eqn:preconditioners}
  d_t( f) :=  f\oslash {A_t^\top{1}}.
\end{equation}
An update of this type is a standard extension of OSEM algorithms to the MAP problem. However, it does not always perform the maximisation of the given objective at each step.
Note that by including a penalty, the non-negativity of the updates is generally not preserved, and
an additional projection step by $P_{\geq0}$ is applied at each iteration. According to the theory for stochastic gradient ascent, the iteration \eqref{eqn:sga} generally does not
converge to the MAP solution, unless a decaying step-size schedule is employed \cite{dPY01}. This is
attributed to the stochasticity of the gradient estimate, and the variance of the estimated
ascent direction can significantly slow down the convergence when the iterates approach the
maximiser.

One powerful idea to reduce the variance of the gradient estimate in SGA is variance reduction. For the constrained MAP problem, we use proximal versions
of SAGA and SVRG \cite{XZ14} to enforce the iterate feasiblity by projection $P_{\geq0}$. Both SAG and SAGA keep a running table of computed gradients of the subobjectives,
and then efficiently estimate the full gradient. By rescaling the full gradient, SAGA
employs unbiased estimates, whereas the SAG estimate is biased (and thus often harder to
analyze). More precisely, SAGA estimates of the full gradient are given by
\begin{align}\label{eqn:saga_update}
\begin{split}
 \gdir^{(k+1)}_{t_k}&=\nabla \Phi_{t_k}( f_{\rm saga}^{(k)}), \text{ and for } t\neq t_k \text{ keep }\gdir^{(k+1)}_{t}=\gdir^{(k)}_{i};\\
 f_{\rm saga}^{(k+1)} &=   f_{\rm saga}^{(k)} +\alpha{ d_{t_k}( f_{\rm saga}^{(k)})\odot} \bigg(\gdir^{(k+1)}_{t_k}- \gdir^{(k)}_{t_k}+\frac{1}{N_s}\sum_{t=1}^{N_s}\gdir^{(k)}_{t}\bigg).
\end{split}
\end{align}

SVRG is an unbiased variance reduction method that has inspired SVREM, and mirrors the convergence rate performance of SAG and SAGA, but does not require maintaining a running list of gradients.
Setting  $ f^{\text{anc}}= f^{(0)}$, and $\tilde{\gdir}= \frac{1}{N_s}\nabla\mathcal{L}( f^{\text{anc}}) -\frac{\beta}{N_s}\nabla\mathcal{R}( f^{\text{anc}})$
the algorithm follows
\begin{align}\label{eqn:svrg_update}
\begin{split}
 f_{\rm svrg}^{(k+1)} &=   f_{\rm svrg}^{(k)} +\alpha  d_{t_k}( f_{\rm svrg}^{(k)})\odot\bigg(\nabla \Phi_{t_k}( f_{\rm svrg}^{(k)}) - \nabla \Phi_{t_k}( f^{\text{anc}})+\tilde{\gdir}\bigg).\\
\text{If } k\,\, {\rm mod}\,\, \ufrq N_s &= 0 \text{ set the anchor estimate }  f^{\text{anc}}=  f_{\rm svrg}^{(k)} \text { and update } \tilde \gdir = \frac{1}{N_s}\nabla \Phi( f^{\text{anc}}).
\end{split}
\end{align}
A value of the full gradient update frequency $\ufrq$ between $2$ and $5$ is recommended \cite{JZ13}.

Several remarks are in order.
First, note that the methods described in Sections \ref{sec:EM} and \ref{sec:regularisation} aim
at explicitly computing the maximiser at each step. However, update
equations \eqref{eqn:saga_update}, and \eqref{eqn:svrg_update} are rather derived by
analogy with the additive formulation \eqref{eqn:osem_additive}, and thus mathematically less principled. Second, provided that
the subobjectives $\Phi_t$ are $L$-Lipschitz, SAG and SAGA{} converge (sub-linearly
in expectation) to the minimiser for the fixed stepsize $\alpha=(16L)^{-1}$ \cite{SRB17}.
This result does not apply to the PET problem \eqref{eqn:PET-MAP} since the subobjectives $\mathcal{L}_t$ are
not Lipschitz in a neighbourhood of $0$, and since the algorithms use iteration-dependent preconditioners. 
Third, as observed in \cite{AF03}, preconditioned gradient ascent based algorithms (which are in PET literature sometimes called diagonally-scaled incremental gradient methods) do
not always converge to the MAP solution when using iteration-dependent preconditioners.
Indeed, assuming $ f^{(k)}\rightarrow  f^\star$, that $\nabla \Phi( f)$ is continuous, and that preconditioner functions $ d_t( f)\geq0$
are continuous such that $\| d_t( f)\|\neq0$ for $ f\neq 0$, the issue seems to persist for the proposed stochastic algorithms. The latter assumption is implicitly satisfied by the EM preconditioner.
Assuming $\lim_{k\rightarrow\infty}\alpha_k=0$ and $\sum_{k=1}^\infty \alpha_k=\infty$, then the
convergence of the SGA{} algorithm implies
\[
\sum_{t=1}^{N_s}  d_t( f^\star)\odot\nabla \Phi_t( f^\star)=0.
\]
Thus, unless all the preconditioners $d_t$ are the same, this identity generally is different from the true optimality condition
$\nabla \Phi( f^\star)=0$ (for unconstrained optimisation). Note that an analogous analysis holds
for stochastic estimators in expectation, provided the corresponding search direction $\sdir_k( f^{(k)},
\xi_k)$ is unbiased and consistent. A simple remedy is to freeze the preconditioner,
$ d_t( f^{(k)}) = d$, where $d$ is constant and given as $d=d(f^{(0)})$. In practice, gradient-based variance reduction methods, with either
iteration dependent or constant preconditioners, perform well for the PET{} problem \cite{TA+20,TAT21}.

The convergence of non-preconditioned SAGA and SVRG has mostly been studied for strongly convex objectives.
Namely, SAGA enjoys linear convergence for strongly convex problems \cite{DBLJ14}, and $
O(1/k)$ convergence of the average iterate for convex problems \cite[Theorem 4.8]{D14}, whereas SVRG converges
at a linear rate for strongly convex problems (evaluated at the anchor point)
\cite{JZ13}. These assumptions are not satisfied by the PET problem, neither for standard
nor modified likelihood \eqref{eqn:modified_LL}, which is only strictly convex.
Nonetheless, combining the arguments in \cite{CLS18} and \cite{AF03} allows
establishing almost sure convergence of both SAGA{} and SVRG for a modified
likelihood term and a constant preconditioner. In practice, this holds since
the background events are nonzero, for which there is no need to modify the likelihood.
\begin{theorem}\label{thm:sagasvrg_convergence}
Let $\widetilde\Phi( f)=\widetilde{\mathcal{L}}( f)-\beta\mathcal{R}( f)$ where $\widetilde{\mathcal{L}}( f)=\sum_{i=1}^M \pll_i(a_i^\top f)$ uses the modified likelihood \eqref{eqn:modified_LL}.
Moreover, let all entries of $ d\in\mathbb{R}^N$ be positive, denote by $L=\max_{t\in N_s} L_t$ the largest of the
Lipschitz constant of sub-objective gradients $\widetilde\Phi_t( f)$ and by $d_{\max}=\|d\|_\infty$ the largest entry
of $ d$, and assume $\argmax_{ f\geq0}\Phi( f)\neq\emptyset$.

Then taking $\alpha=\frac{1}{3Ld_{\max}^{1/2}}$ and $ d_t( f_{\rm saga}^{(k)})= d$ in the SAGA{} algorithm \eqref{eqn:saga_update}
we have $\widetilde\Phi( f_{\rm saga}^{(k)})\rightarrow\Phi( f^\star)$ and $ f_{\rm saga}^{(k)}\rightarrow f^\star$ almost surely.
Taking $\alpha\leq\frac{1}{4Ld_{\max}^{1/2}{(\ufrq N_s+2)}}$ and $ d_t( f_{\rm svrg}^{(k)})= d$ in the {\rm svrg}{} algorithm \eqref{eqn:svrg_update} we have
$ f_{\rm svrg}^{(k)}\rightarrow f^\star$ almost surely and $\mathbb{E}[\Phi( f^\star)-\widetilde\Phi( f_{\rm svrg}^{(k\ufrq N_s)})]=\mathcal{O}(1/k)$.
\end{theorem}

Note that Theorem \ref{thm:sagasvrg_convergence} provides conditions for convergence but does not quantify the speed of convergence. This is typical for non-strongly concave problems for most variance reduction techniques.
We can draw an analogy with the results for strongly-concave problems which suggest that the speed of convergence depends on the ratio between the Lipschitz constant and the stepsize.
In case of preconditioned SVRG{} and SAGA the effective stepsize is determined not only by $\alpha$ but also by the preconditioner $ d$, as indicated by the bounds on the stepsize in Theorem \ref{thm:sagasvrg_convergence}.

\section{Numerical experiments and discussions}\label{sec:numerics}
This section presents numerical results for the two classes of stochastic methods for the MAP problem
\eqref{eqn:PET-MAP} with the $\log \cosh$ penalty, which is often employed for PET reconstruction.
We present two examples: a brain phantom, and a torso XCAT.
In these examples we examine the performance of SVREM with SVRG and SAGA.
SVRG and SAGA are employed with a constant preconditioner $ d= d( f^{(0)})$,
which provides consistently better performance than the iteration dependent counterpart $d(f^{(k)})$. The results for SAG and SARAH are
nearly identical with those for SAGA and SVRG, respectively, and thus are not included.

\subsection{Brain phantom}
In this experiment, we take a single  slice (of size $114\times114$) from a brain phantom, available at \url{https://github.com/casperdcl/brainweb}. The forward map is taken to be the Radon transform
using $180$ projection angles with a $1$ angle separation. The sinogram data was binned into
subsets using geometric projections of the scanner.
We consider $N_s=15$, $30$, and $45$ subsets and accordingly, each subset consists of $12, 6$, and $4$ views, respectively. In the reconstruction,
we use the $\log\cosh$ penalty with $\delta =0.01$ and a regularisation parameter
$\beta = 60$, which are determined in a trial-and-error manner, and conduct the experiments in \texttt{MATLAB R2019b}. The accuracy of a reconstruction
$f$ is measured by the relative error $\Delta( f) = \|{ f- f^\star}\|/{\|{ f^\star}\|},$
where $ f^\star$ is the reference solution, computed using LBFGS-B
(available at \url{https://www.mathworks.com/matlabcentral/fileexchange/35104-lbfgsb-l-bfgs-b-mex-wrapper},
retrieved on April 11, 2021).
All algorithms are initialised with one epoch of OSEM.
Following \cite{TA+20}, for SAGA and SVRG, we employ a constant stepsize $\alpha=2$, and update SVRG with the full 
gradient every $\ufrq=2$ epochs, whereas for SVREM, we choose $\alpha = 0.7$ and $\ufrq=1$. We consider three OSEM type methods without variance reduction, i.e., BSREM, SGA, and SEM, which are representative within PET reconstruction, as baselines. More precisely, SGA is the PET reconstruction method, described in \eqref{eqn:sga}, which for the full objective $\Phi$ reads
\begin{equation}
 f_{\rm sga}^{(k+1)} =  f_{\rm sga}^{(k)} + \alpha({ f_{\rm sga}^{(k)}}\oslash {A_{t_k}^\top {1}})\odot\nabla \Phi_{t_k}( f_{\rm sga}^{(k)}).
\end{equation}
BSREM \cite{AF03} is an iterative scheme based on OSEM that uses a decaying stepsize schedule, i.e.,
\begin{equation}
 f_{\rm bsrem}^{(k+1)} =  f_{\rm bsrem}^{(k)} + \alpha_k({ f_{\rm bsrem}^{(k)}}\oslash {A^\top {1}})\odot\nabla \Phi_{t_k}( f_{\rm bsrem}^{(k)}).
\end{equation}
Note that standard BSREM uses a subset-independent preconditioner. We use its subset-dependent variant, replacing $A^\top 1$  with $A_{t_k}^\top 1$ since it exhibits a faster, yet steady, convergence in the studied setting.
In the experiments, both SEM (cf. \eqref{eqn:tempsem}) and BSREM use the stepsize schedule $\alpha_k=(0.001k+1)^{-1}$, which is sufficient to ensure their convergence. Since SVRG and SVREM require the full gradient once every 
$\ufrq$ epochs, to make a fair comparison of overall computational cost, we count epochs in terms of the number of subsets that are
 used at each iteration. Thus, every $\ufrq N_s$ updates of SVRG and SVREM are counted as $\ufrq+1$ epochs, and $\ufrq$ 
epochs of SAGA. 
 
\begin{figure}[h!]
\centering
\setlength{\tabcolsep}{0pt}
\begin{tabular}{cc}
\includegraphics[width=.478\textwidth]{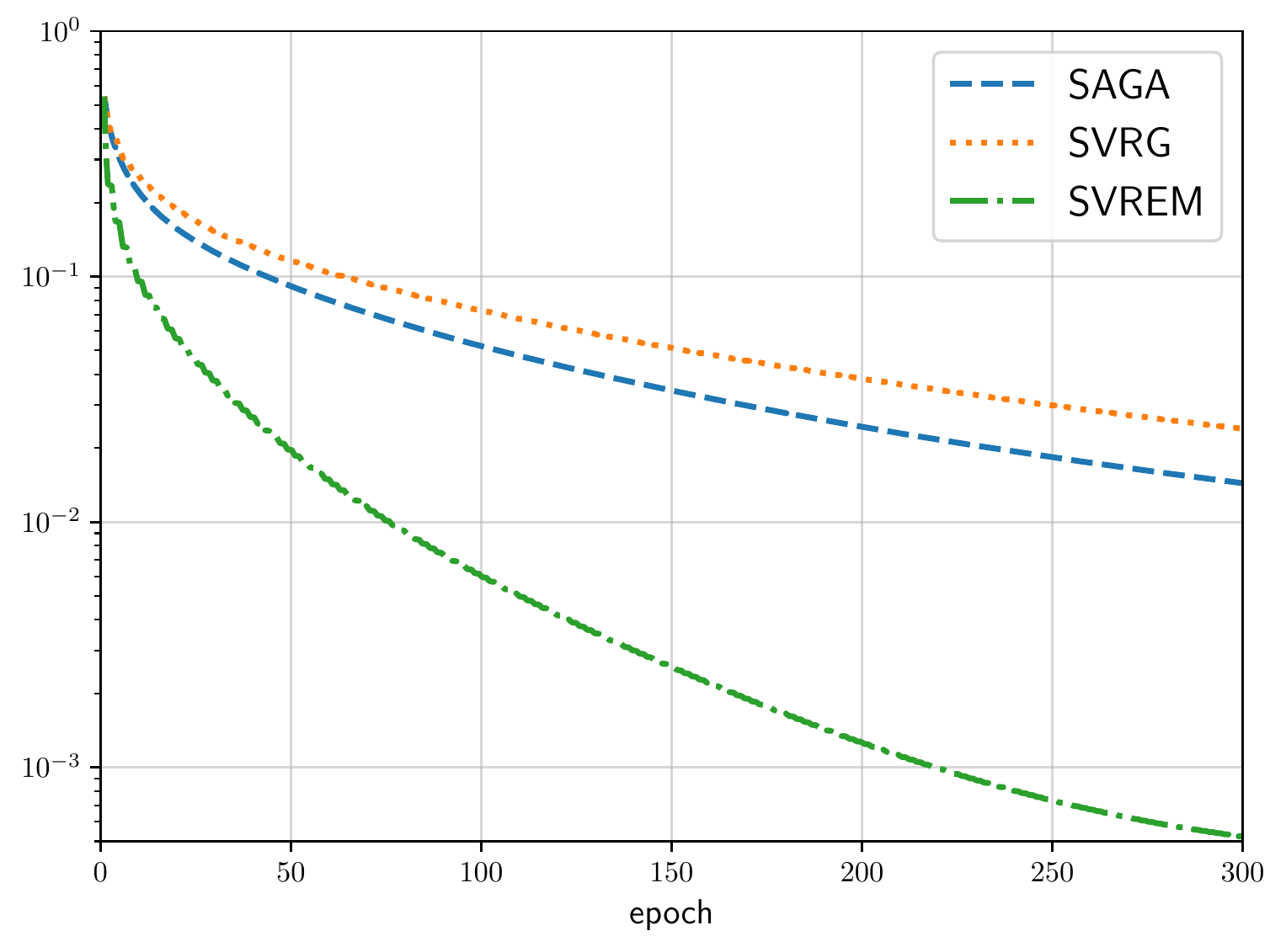}
& \includegraphics[width=.478\textwidth]{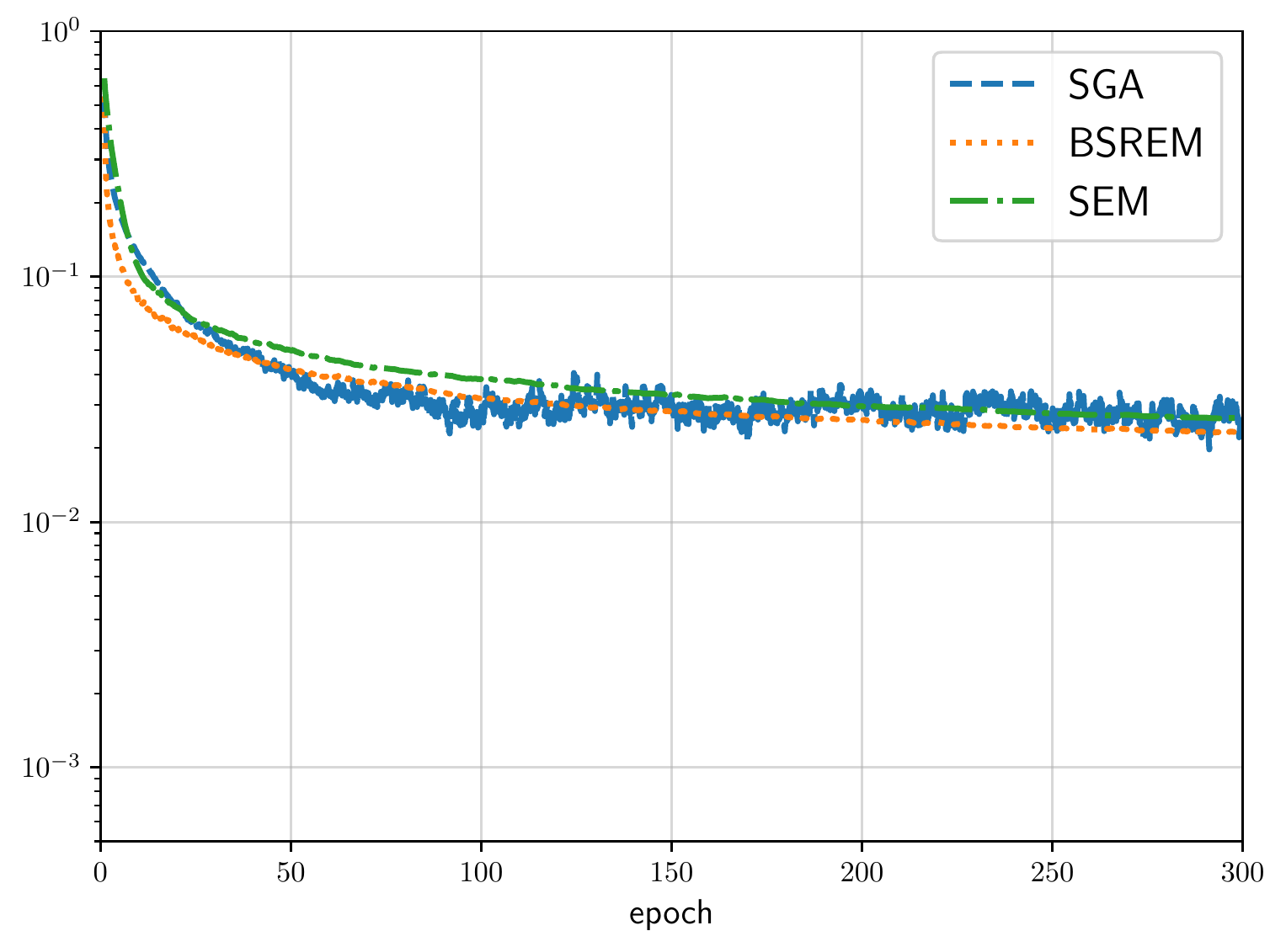} \\
(a) Variance reduction & (b) Non-Variance reduction methods
\end{tabular}
\caption{The error decay behaviour for the brain phantom: (a) SAGA, SVRG, and SVREM, and (b) OSEM, BSREM, and SEM. 
}
\label{fig:LOGCH_Brain_Delta}
\end{figure}

\begin{figure}[h!]
\centering
\setlength{\tabcolsep}{0pt}
\begin{tabular}{cccc}
\includegraphics[width=.24\textwidth]{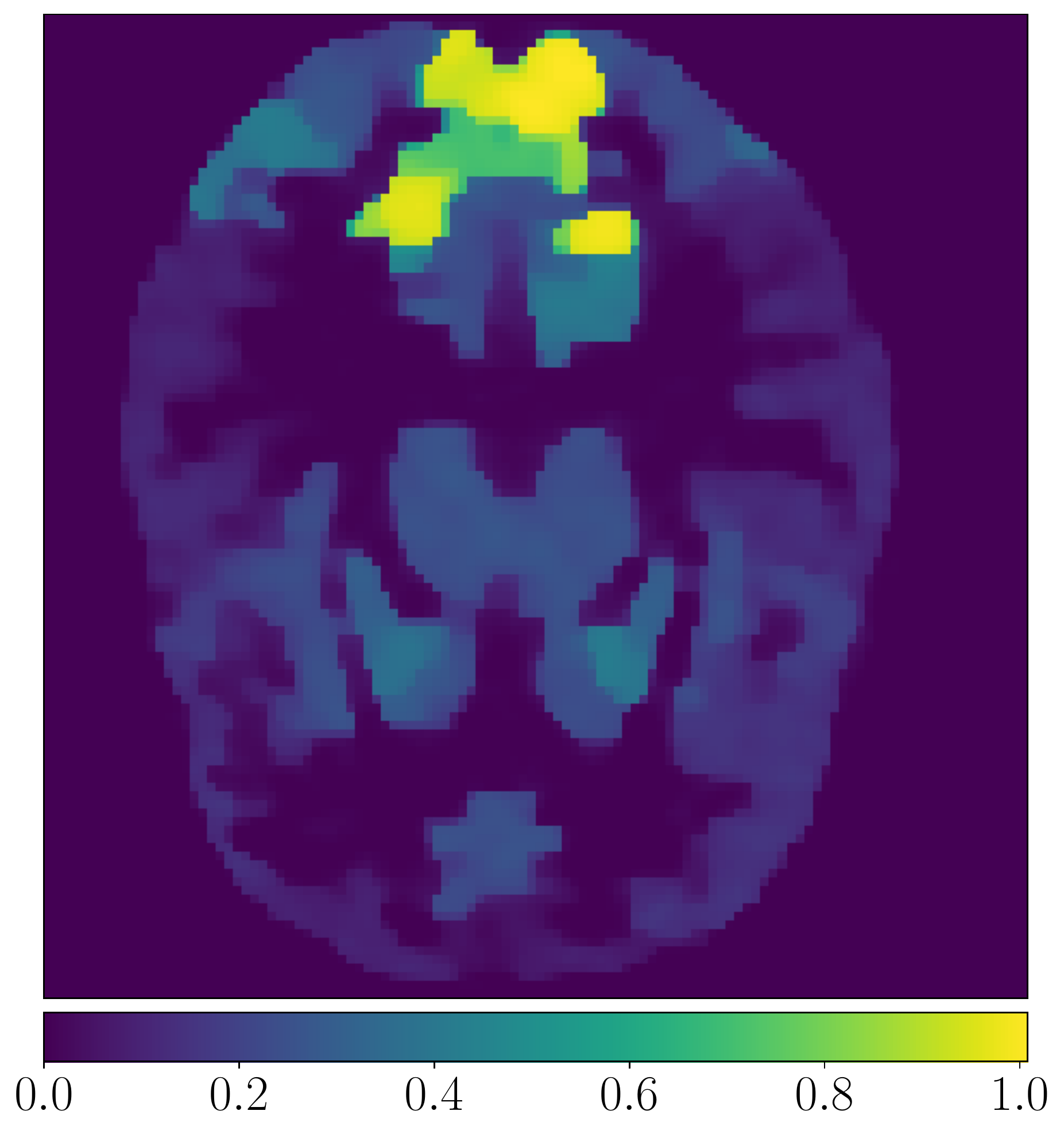} &  \includegraphics[width=.25\textwidth]{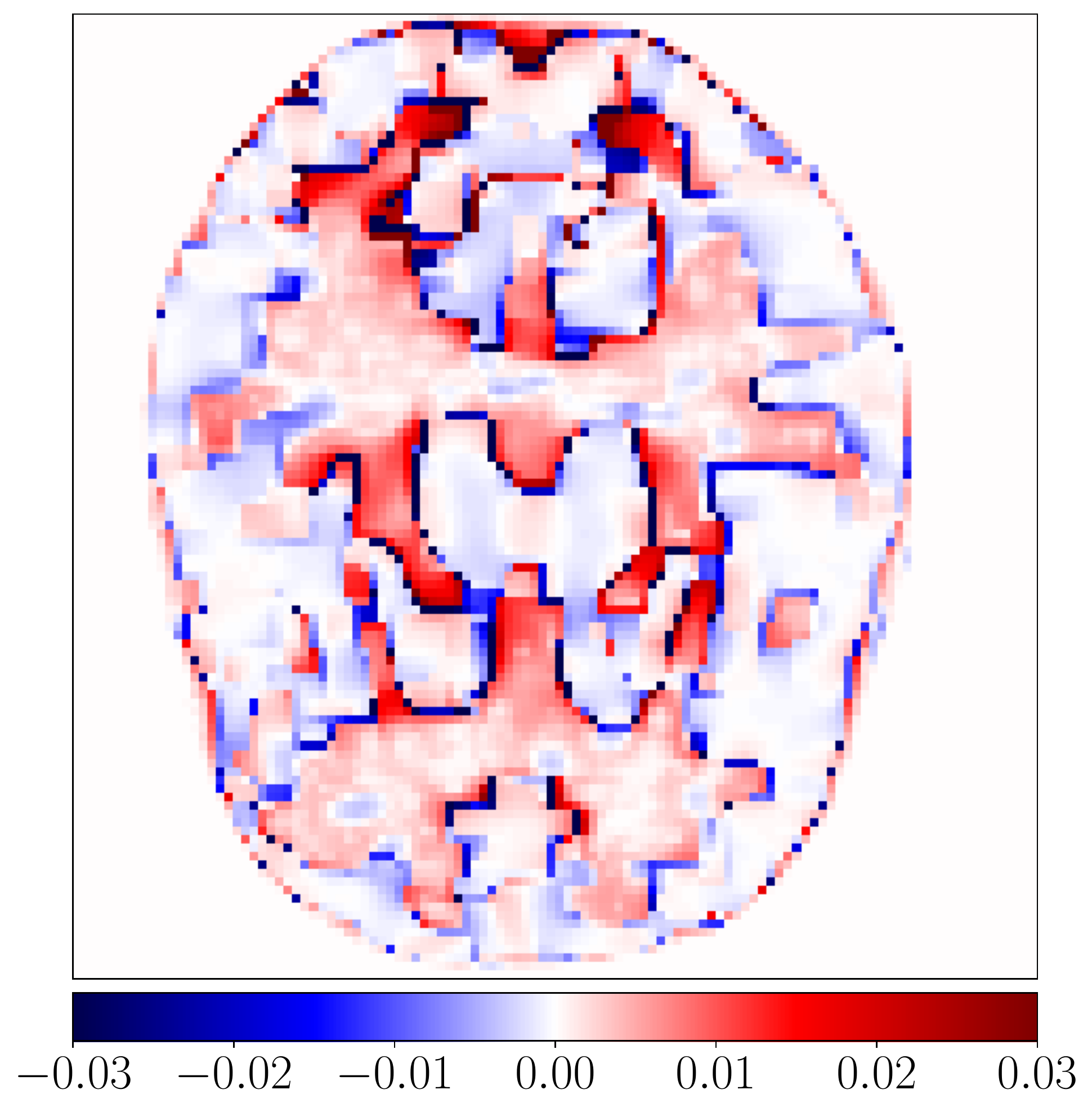} & \includegraphics[width=.25\textwidth]{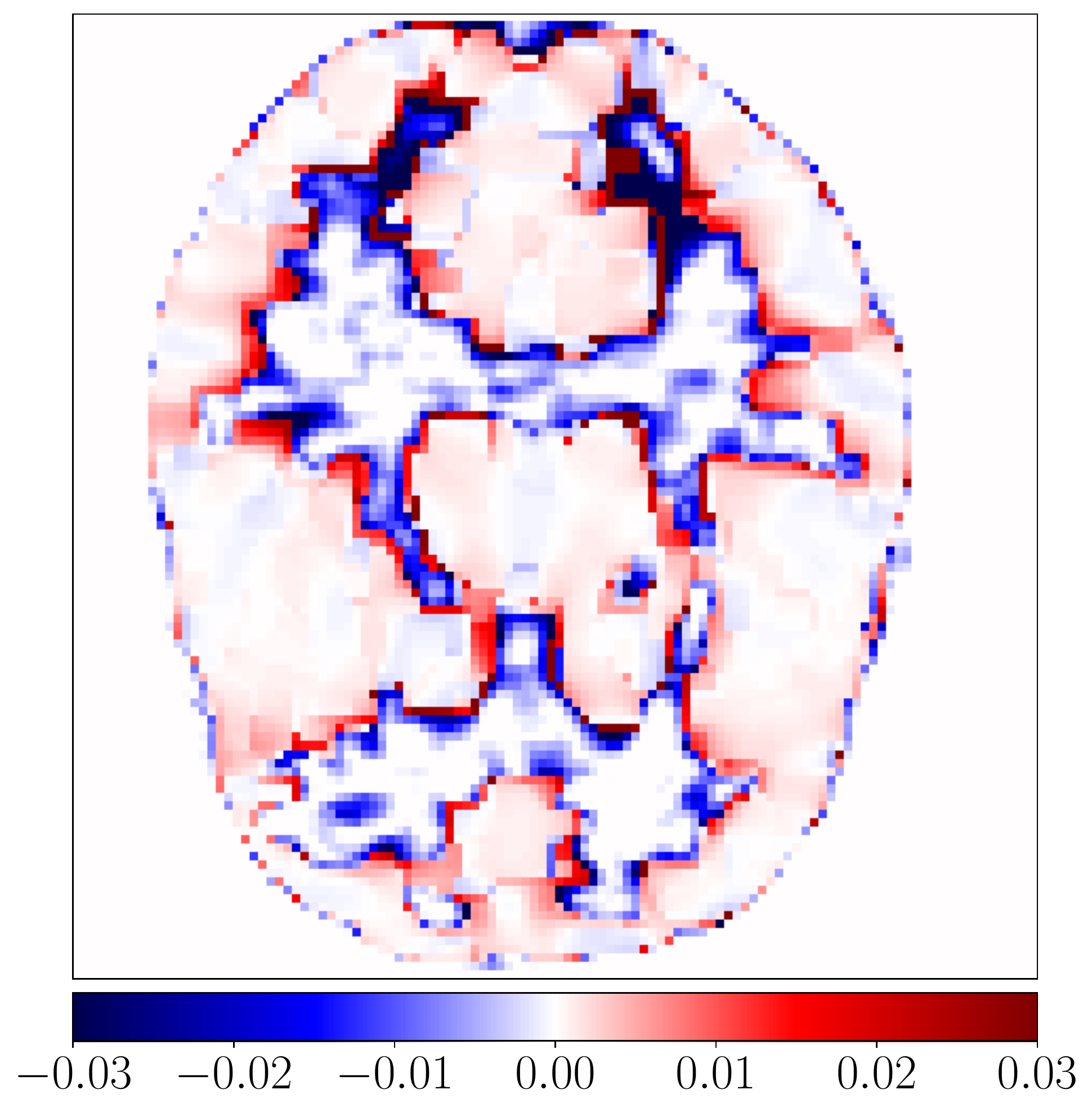} & \includegraphics[width=.25\textwidth]{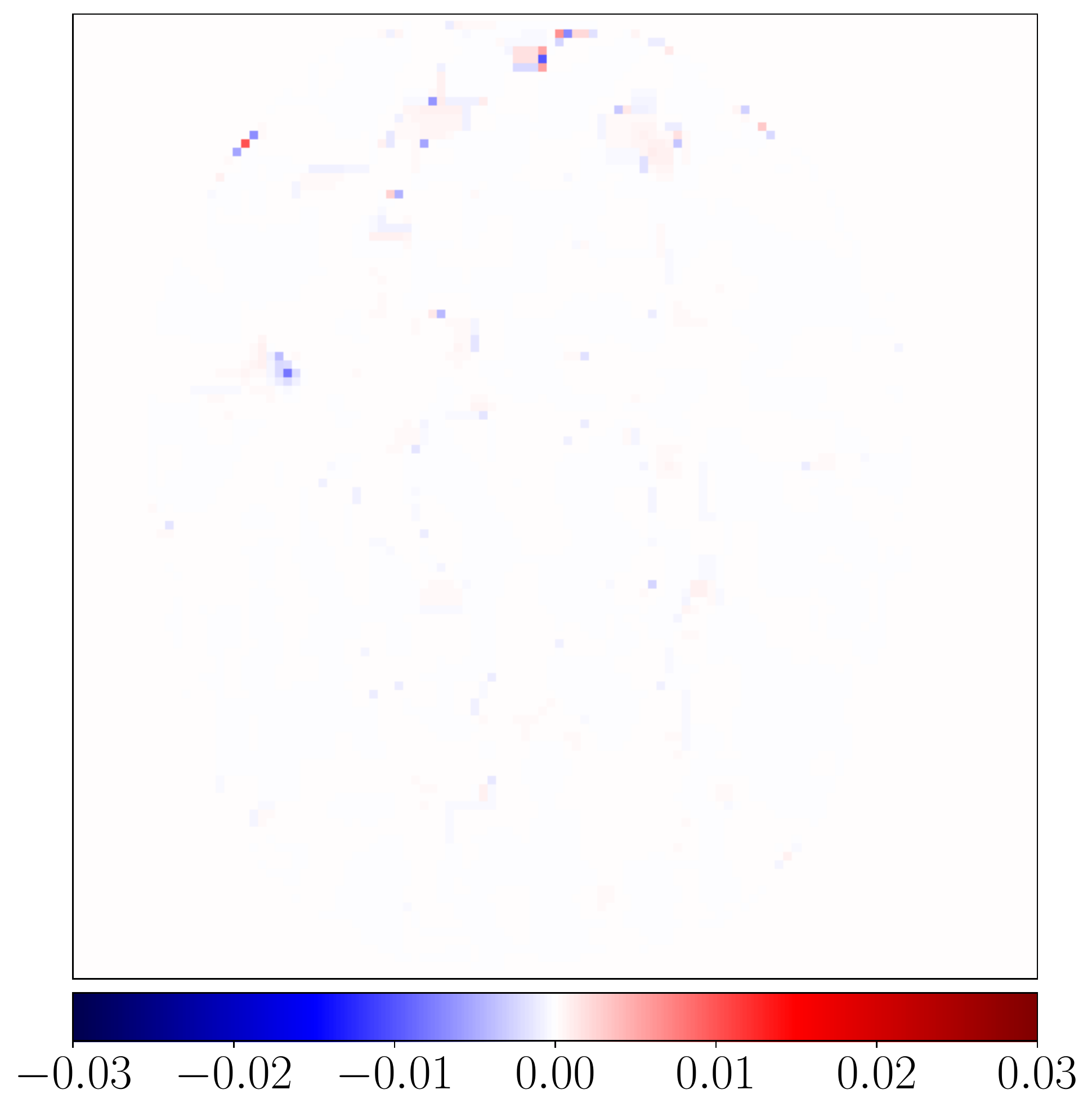}\\
(a) reference & (b) SAGA & (c) SVRG & (d) SVREM
\end{tabular}
\caption{The pixel wise errors of SVRG, SAGA, and SVREM{} for brain phantom with respect to the reference solution computed by LBFGS-B.}
\label{fig:LOGCH_Brain_PixelErrors}
\end{figure}

In Fig.~\ref{fig:LOGCH_Brain_Delta}, we show the comparative results for all methods on $N_s=30$ subsets of the data. It is clearly observed that SVREM exhibits the fastest convergence among all the methods, which is also corroborated by the pixel-wise errors in Fig.~\ref{fig:LOGCH_Brain_PixelErrors}: the SVREM reconstruction agrees nearly perfectly with the 
reference solution, whereas the pixelwise errors of SAGA and SVRG reconstructions still clearly exhibit structures, especially edges. Moreover, SAGA is slightly slower than SVRG (in terms of the running time), and both algorithms would benefit from a larger stepsize, though setting it too large impacts the overall convergence. We will explore the stepsize issue for SAGA and SVRG in Section \ref{subsec:torso}.
Just as expected, all variance reduction methods outperform OSEM, BSREM, and SEM, and are orders of magnitude faster, especially for high-accuracy solutions, showing clearly the beneficial effect of variance reduction for accelerating OSEM type algorithms. Although not presented, one can observe similar behavior for other subset numbers.

\begin{figure}[h!]
\centering
\setlength{\tabcolsep}{2pt}
\begin{tabular}{ccc}
\includegraphics[width=.33\textwidth]{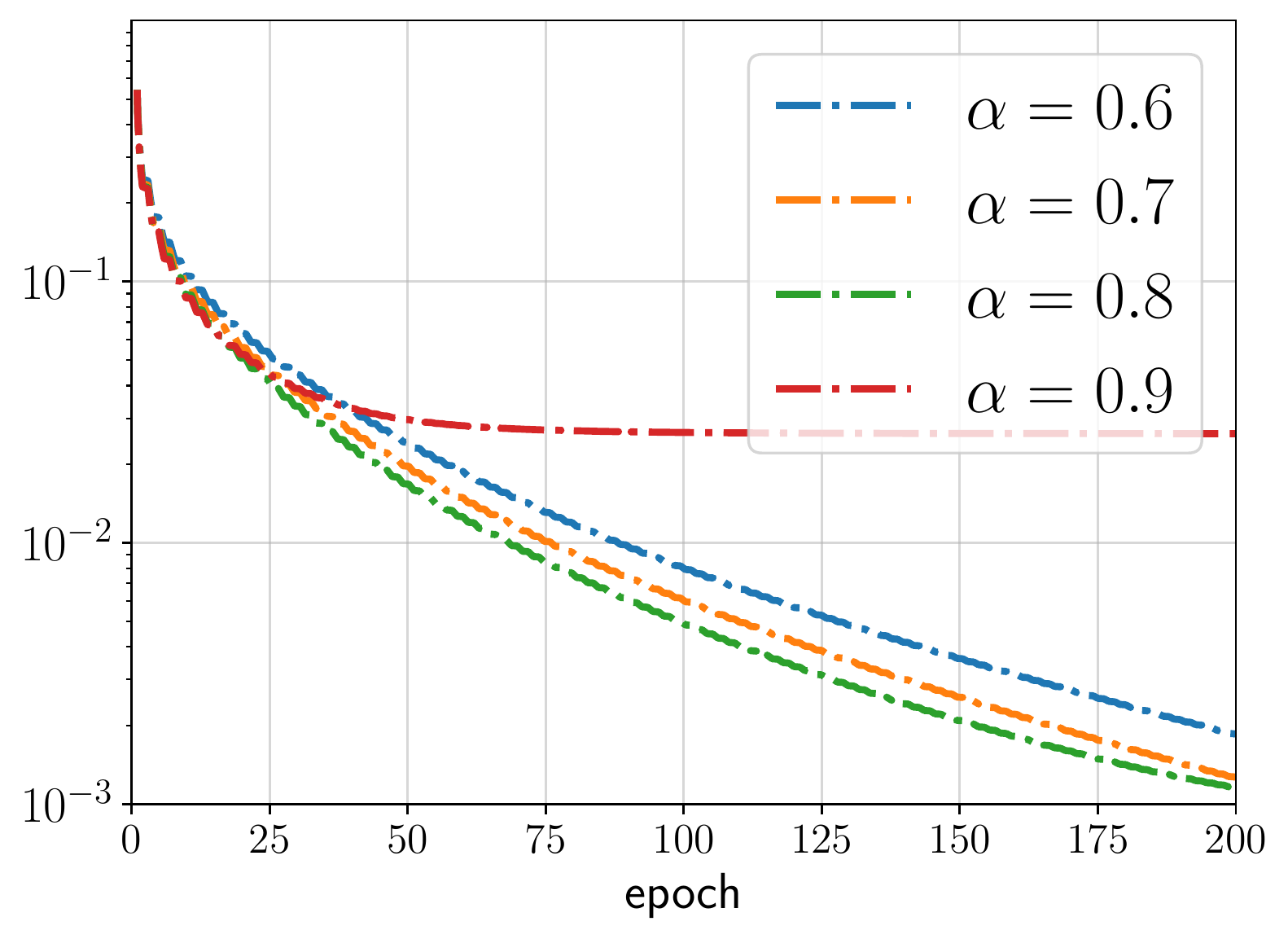}& \includegraphics[width=.33\textwidth]{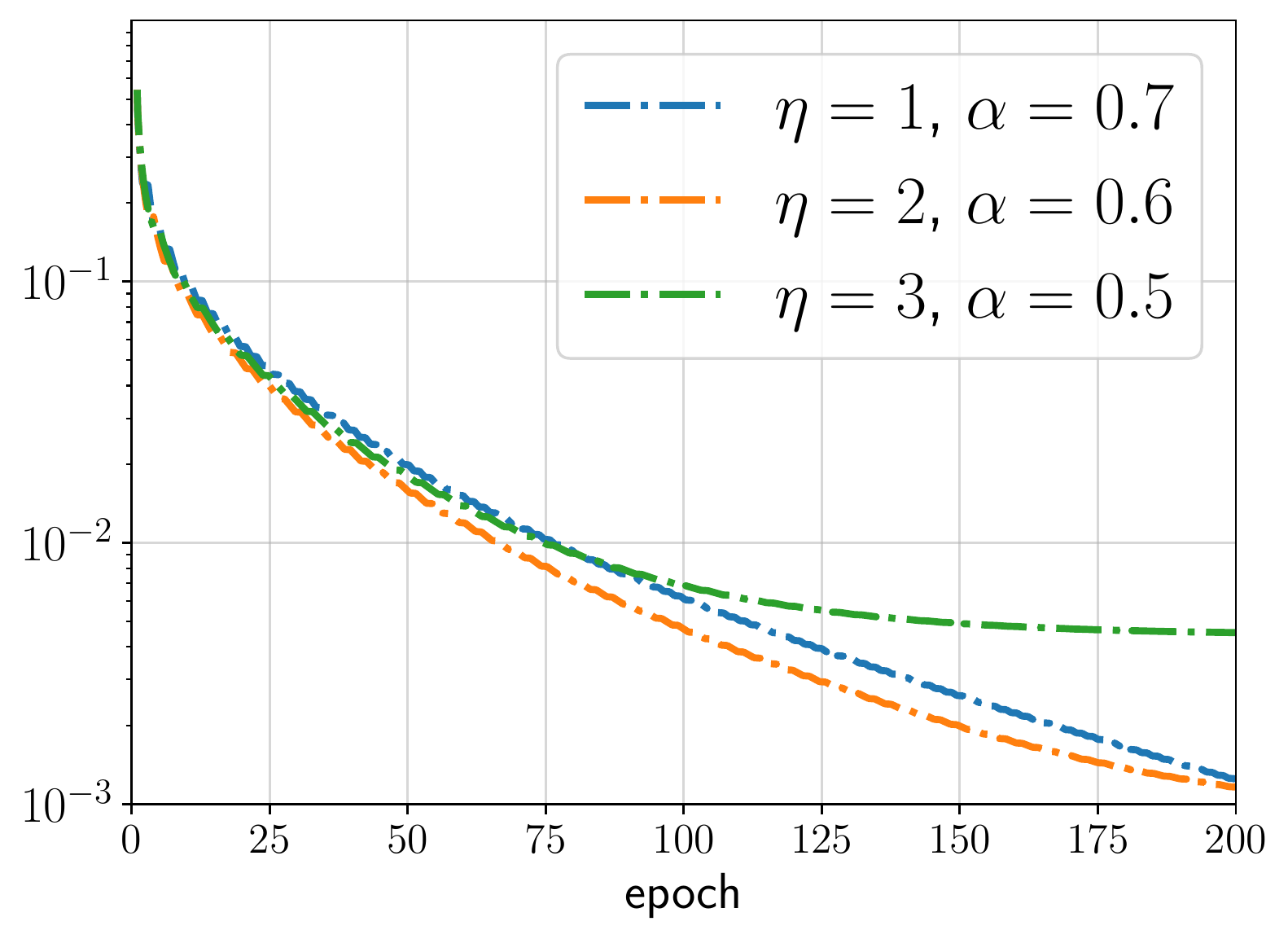} & \includegraphics[width=.33\textwidth]{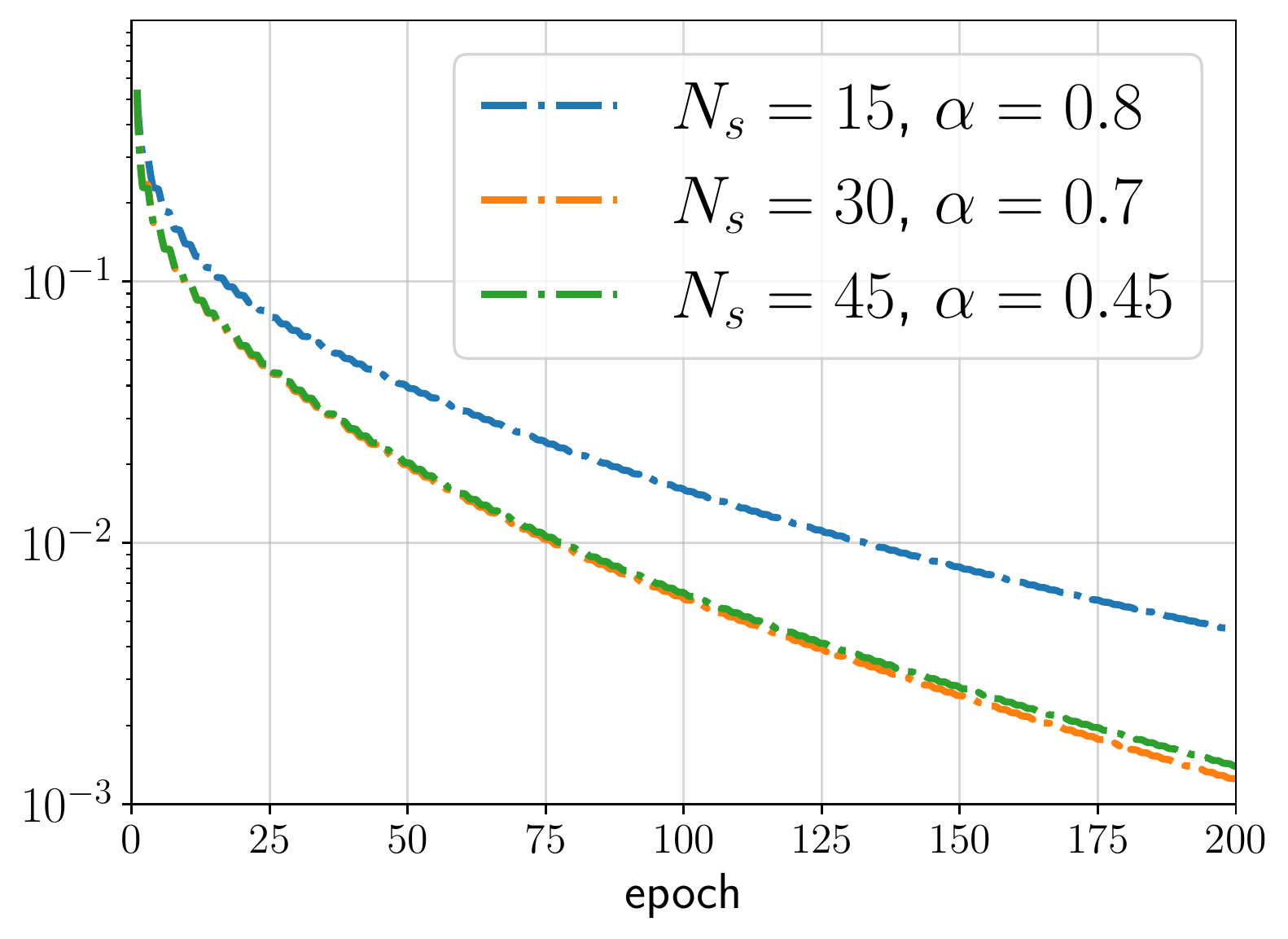} \\
(a) Stepsizes & (b) Update frequency & (c) Number of subsets
\end{tabular}
\caption{The error decay behaviour for SVREM on the brain phantom with respect to the stepsize $\alpha$, the update frequency $\zeta$ the number $N_s$ of subsets.  
In panels (a) and (b) we use $30$ subsets of the data.
}
\label{fig:LOGCH_Brain_SVREM}
\end{figure}

Fig.~\ref{fig:LOGCH_Brain_SVREM} studies the convergence behavior of SVREM with respect to three important algorithmic 
parameters, i.e., update frequency $\ufrq$ (of full gradient updates), the stepsize $\alpha$, and number 
of subsets $N_s$. Generally, all these parameters greatly impact the performance of SVREM, and they should be tuned simultaneously to achieve optimal convergence behavior. 
It is observed that increasing the 
frequency $\ufrq$ provides some acceleration in initial epochs but a too large $\ufrq$
can impair the asymptotic convergence, even with a tuned stepsize $\alpha$. The stepsize $\alpha$
has a similar influence as the frequency $\ufrq$: a larger stepsize $\alpha$ gives faster
initial acceleration, but too large a value may prevent the algorithm from converging to the MAP maximiser. Lastly, a larger number $N_s$ of subsets (with suitably tuned stepsizes) tends to provide faster initial  convergence. This observation shows clearly the importance of proper partition of the subsets.

\subsection{Torso phantom}\label{subsec:torso}

To evaluate the algorithms in a more realistic PET setting, in this experiment we use a PET{} scan of a torso, obtained as an XCAT simulated phantom \cite{Segars2010} with 2 rings and 280 projection angles.
The sinogram data thus consists of $280$
views, and was binned into $20, 40$, and $70$ subsets using geometric projections of the scanner, so that each subset consists
of $14$, $7$, and $4$ views, respectively. In the experiment, SVRG{} and SVREM{} update the full gradient and full
expected statistic, once every $\ufrq=5$ and $\ufrq=3$ epochs, respectively, and all algorithms are initialised with one
epoch of OSEM. The reconstruction was carried out using the well-established Software for Tomographic Image Reconstruction
\cite{STIR}, via a python environment, available at \url{https://github.com/UCL/STIR}.  We use the $\log\cosh$ penalty with $\delta =1$ and fix $\beta = 0.0001$.
Since a reference solution is unavailable, we evaluate the accuracy by the objective value.

\begin{figure}[h!]
\centering
\setlength{\tabcolsep}{0pt}
\begin{tabular}{ccc}
\includegraphics[width=.33\textwidth]{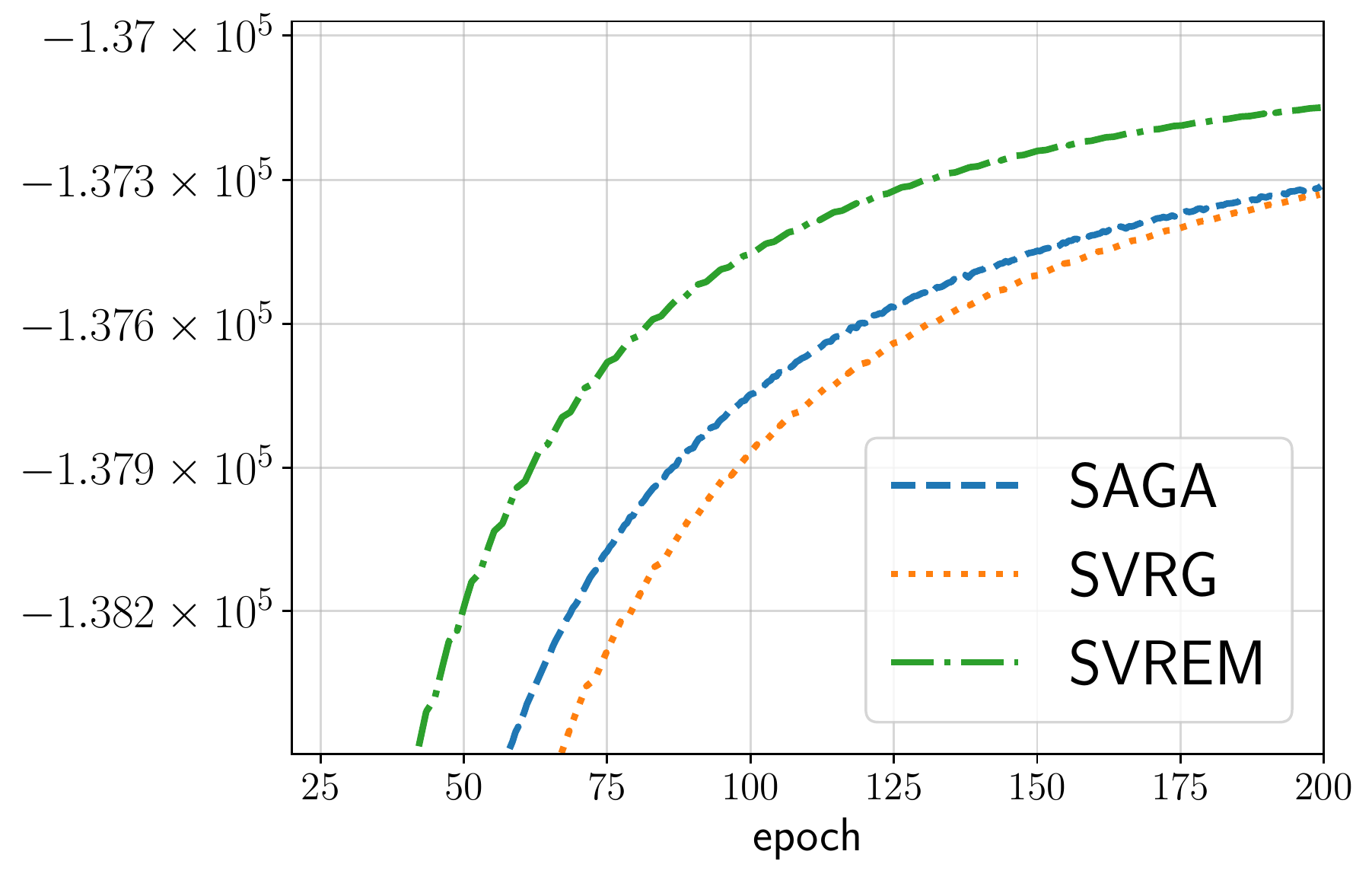} & \includegraphics[width=.33\textwidth]{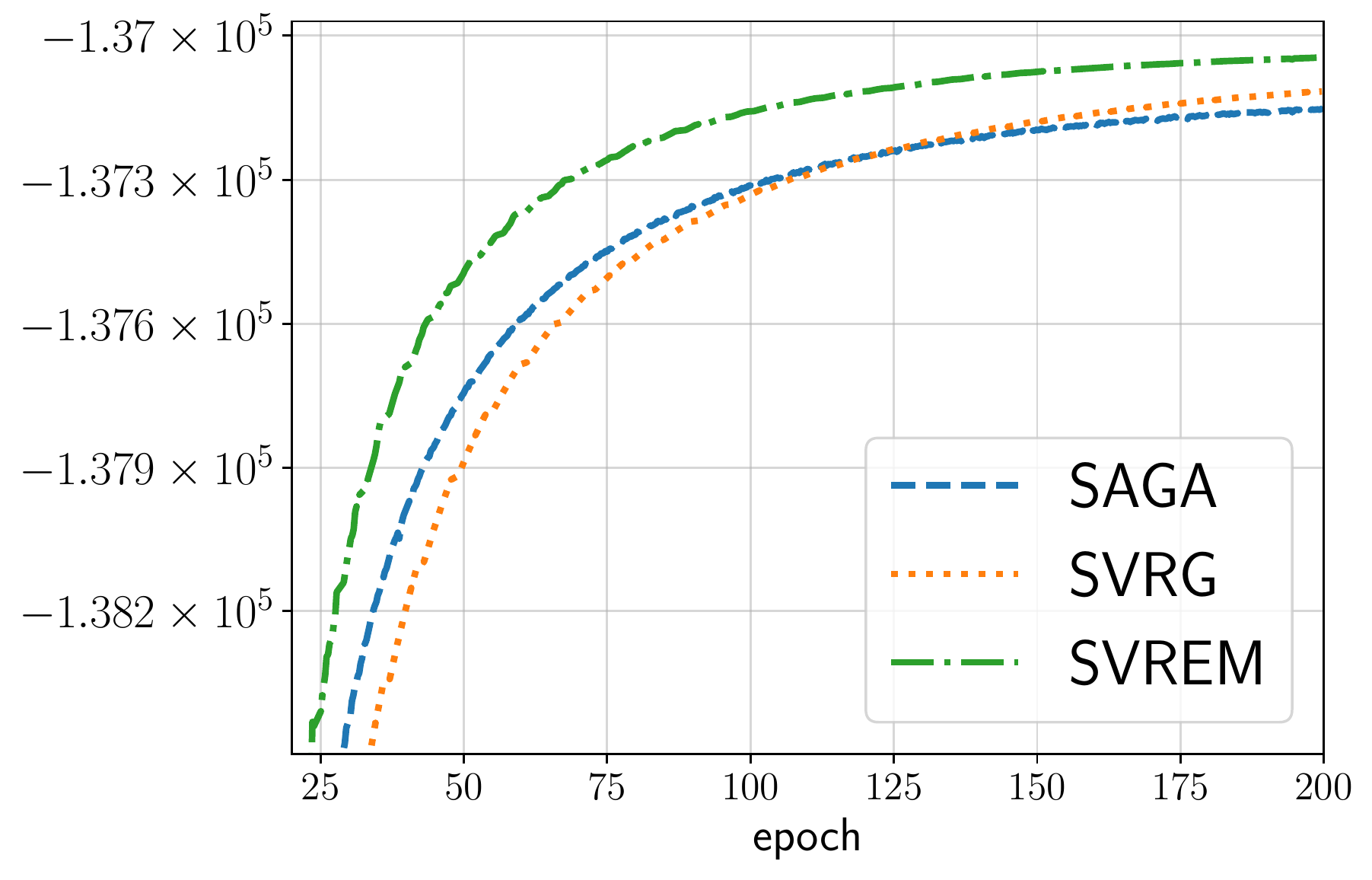} & \includegraphics[width=.33\textwidth]{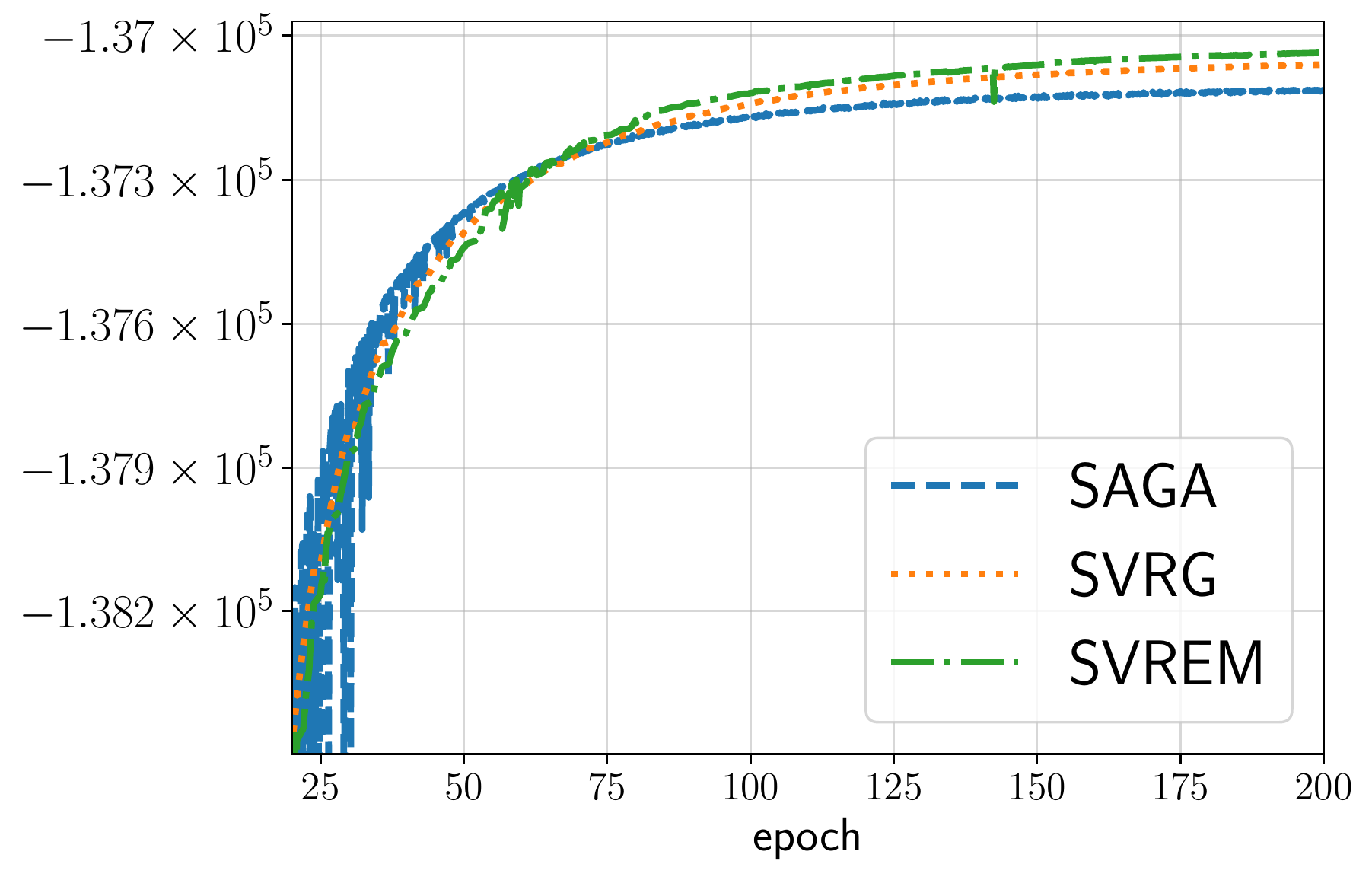} \\
(a) 20 subsets & (b) 40 subsets & (c) 70 subsets
\end{tabular}
\caption{Performance comparison of SAGA, SVRG, and SVREM{} for $20, 40$, and $70$ subsets over $200$ epochs.}
\label{fig:LOGCH_Torso_Delta}
\end{figure}

The results for $200$ epochs of SAGA, SVRG{} and SVREM{} are shown in Fig.~\ref{fig:LOGCH_Torso_Delta}.
Unlike the brain phantom, the behaviour with respect to $\ufrq$ is more stable, and choosing a larger $\ufrq$
provides a better per-epoch comparison with SAGA{}. One interesting observation is about the feasible stepsize regime for SVREM{}.
The work \cite{CZ+18} suggests $\alpha\in(0,1)$, with an upper bound depending on the Lipschitz constant
of subset statistics (and also the number of subsets). This behaviour was also observed in Fig.~\ref{fig:LOGCH_Brain_SVREM}(a).
In contrast, for the torso phantom, the admissible stepsize seem to be $(0,2]$, which corresponds to over-relaxation, a well-known practice in iterative linear solvers \cite{Varga:2000} (see \cite{Tsai:2015} for an application in EM algorithm).
In Fig.~\ref{fig:LOGCH_Torso_Delta}, we use $\alpha=2.0$ for $20$ and $40$ subsets, and $\alpha=1.4$ for $70$ subsets.
In this challenging setting, SAGA, SVRG and SVREM show comparable overall performance, cf. Fig.~\ref{fig:LOGCH_Torso_Delta}.
Nonetheless, SVREM consistently outperforms SVRG, especially when the number $N_s$ of subsets is small, and SAGA, whereas SAGA occasionally exhibits an undesirable stochastic
behaviour during initial iterations (most notably in the case of $70$ subsets), before stabilising after roughly $70$ epochs. The latter can be remedied by choosing a smaller stepsize $\alpha$, which of course can adversely affect the overall convergence behavior. These results suggest 
that SVREM provides a greater benefit for a smaller number $N_s$ of subsets, and as the number $N_s$ of subsets increases 
the algorithms become comparable. 

\begin{figure}[h!]
\centering
\setlength{\tabcolsep}{0pt}
\begin{tabular}{ccc}
\includegraphics[width=.31\textwidth]{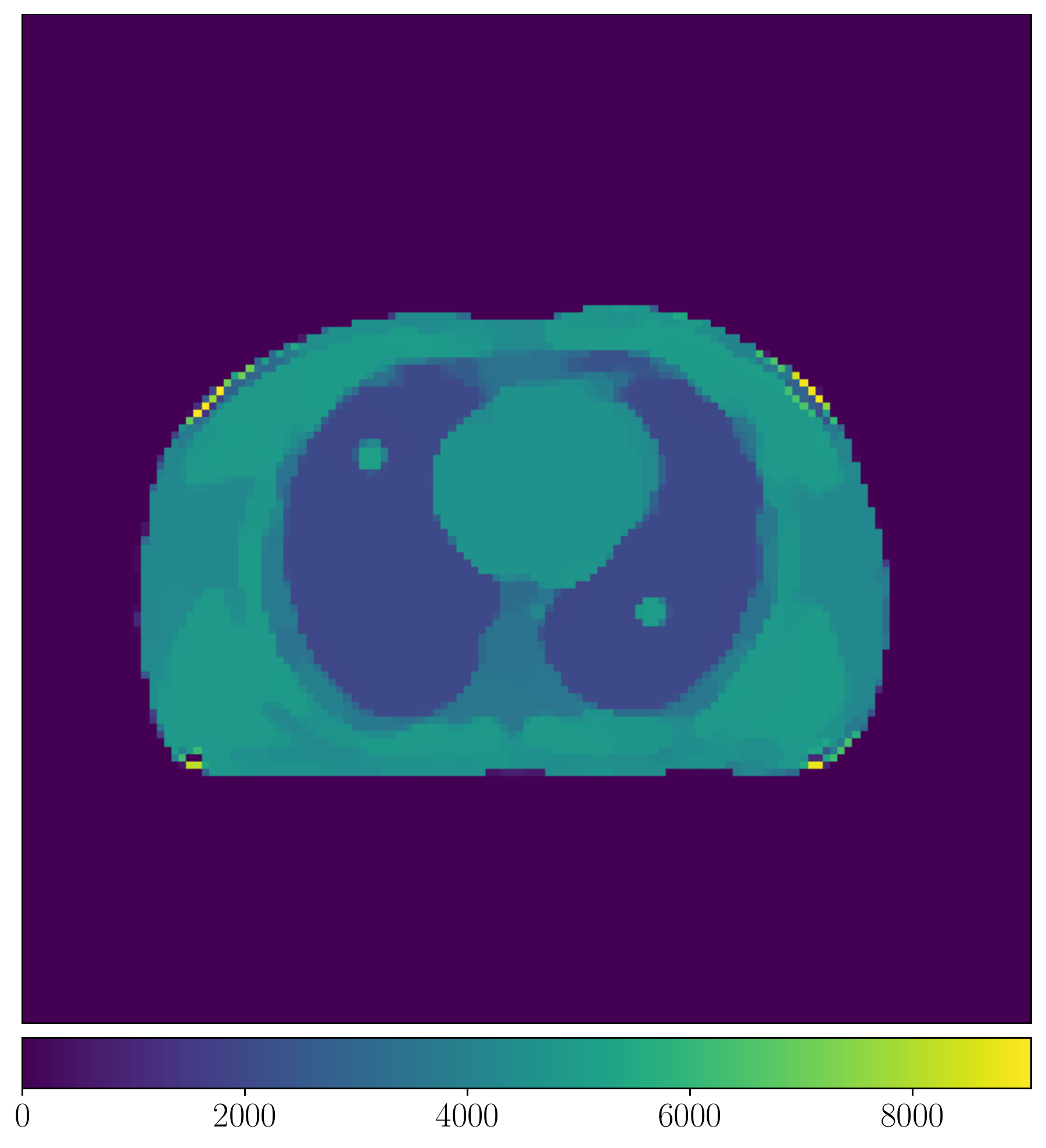} & \includegraphics[width=.33\textwidth]{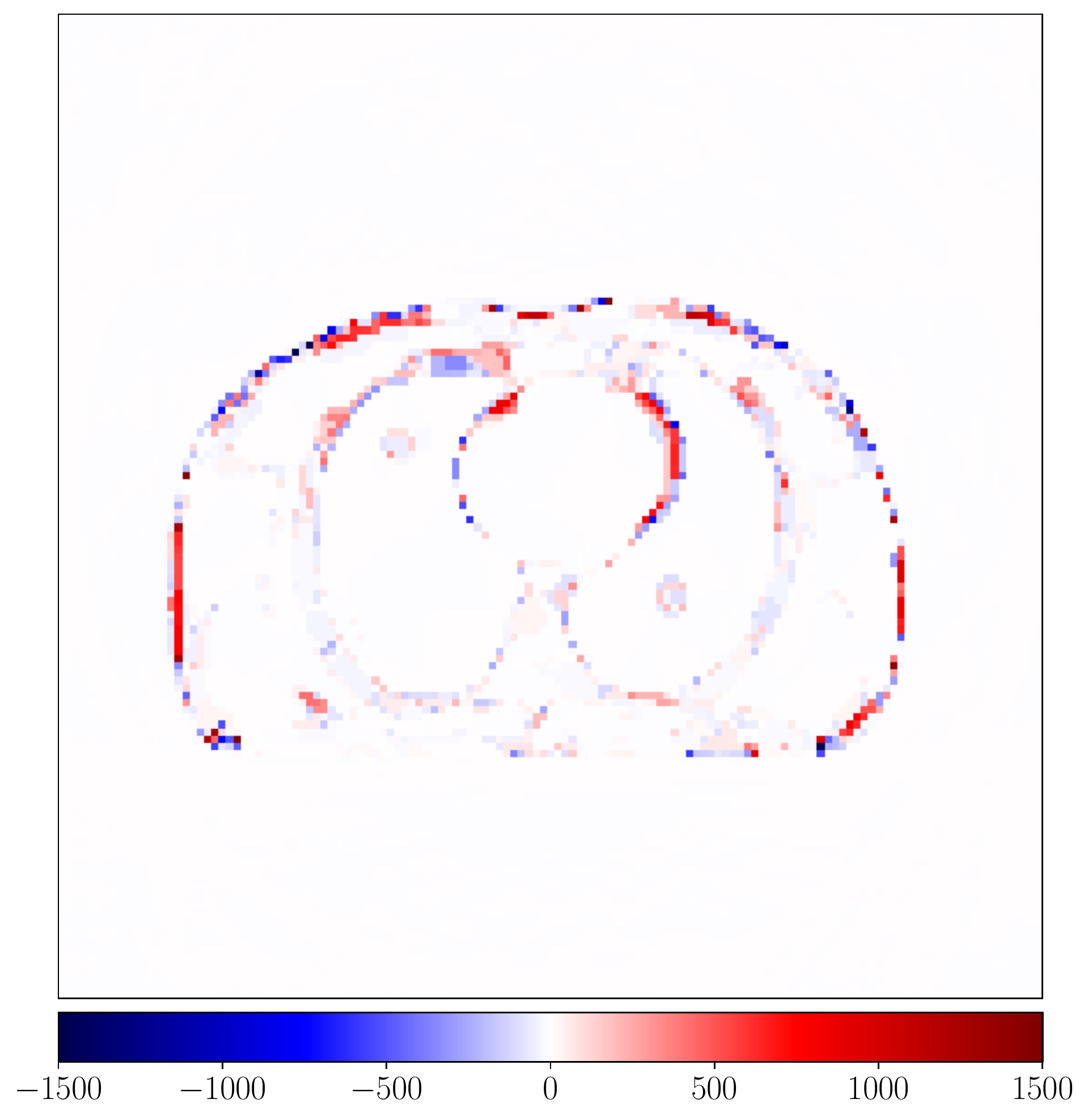} & \includegraphics[width=.33\textwidth]{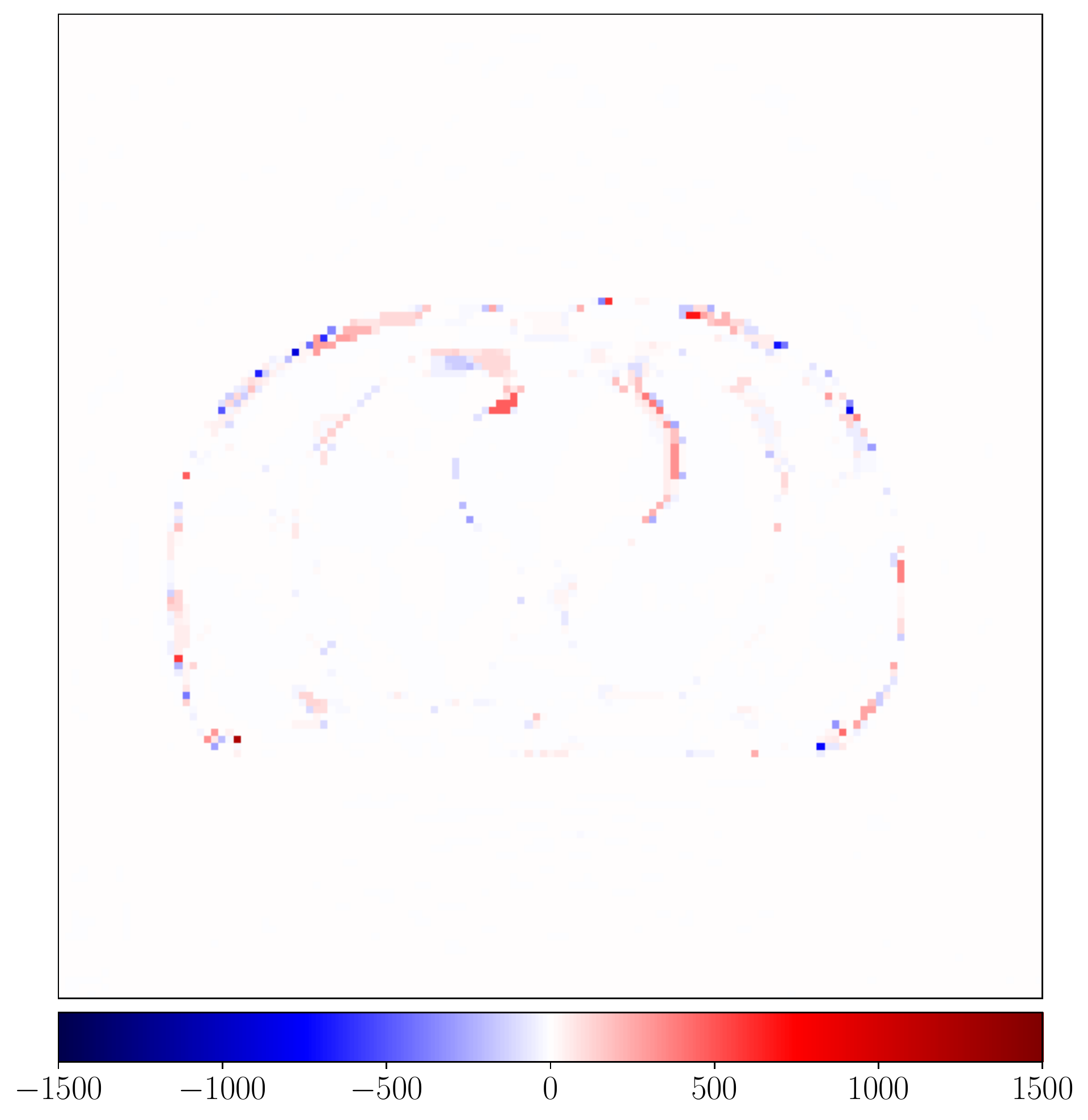}\\
(a) $f_{\rm svrem}^{1000}$ & (b) $ f_{\mathrm{svrem}}^{1000}- f_{\mathrm{svrem}}^{50}$ &  (c) $ f_{\mathrm{svrem}}^{1000}- f_{\mathrm{svrem}}^{200}$
\end{tabular}
\caption{(a) SVREM reconstruction after $1000$, and (b)-(c) pixel-wise differences of SVREM reconstructions after $200$ and $50$ epochs.
}\label{fig:SVREM}
\end{figure}

In Fig.~\ref{fig:Torso-longtime} we study the convergence behaviour over a longer epoch horizon, 
using $40$ subsets and $1000$ epochs, and examine how do SAGA, SVRG, and SVREM depend on the stepsize $\alpha$. The numerical results show that SVREM with $\alpha=1.0$ (and also for smaller values) can eventually outperform both SVRG and SAGA, and  catches up to SVREM with a larger stepsize, confirming the asymptotic convergence.
Meanwhile the results for SVRG, and particularly SAGA, suggest that a smaller stepsize $\alpha$ is needed to ensure their convergence, since otherwise the iterations enter a limit cycle.
Thus, reducing the stepsize has a conflicting effect: it slows down the speed of convergence during initial iterations, but it provides better asymptotic behaviour. 
For a more effective profile of the overall convergence, a dynamically variable stepsize schedule emerges as a natural choice. To gain further insights, in Fig.~\ref{fig:SVREM} we show pixel-wise differences for SVREM with $40$ subsets.
The results show that the background and the smooth parts of the image are mostly 
resolved after $50$ epochs, whereas the edges are more challenging to resolve and are still improving after 50 epochs.

\newcommand{\showpic}[1]{%
\begin{tikzpicture}[spy using outlines={rounded rectangle, magnification=3, width = 2cm, height = 1cm,
connect spies, red, thick}]%
\draw (0,0) node [anchor=north] {\includegraphics[width=.5\textwidth]{#1}};%
\spy on (3.50cm,-0.75cm) in node at (-.2cm, -4cm);
\end{tikzpicture}\hspace*{-2mm}%
}
\newcommand{\showpik}[1]{%
\begin{tikzpicture}[spy using outlines={rounded rectangle, magnification=3, width = 2cm, height = 1cm,
connect spies, red, thick}]%
\draw (0,0) node [anchor=north] {\includegraphics[width=.5\textwidth]{#1}};%
\spy on (3.50cm,-0.52cm) in node at (-.2cm, -4cm);
\end{tikzpicture}\hspace*{-2mm}%
}
\begin{figure}[h!]
\centering
\begin{tabular}{cc}
\showpic{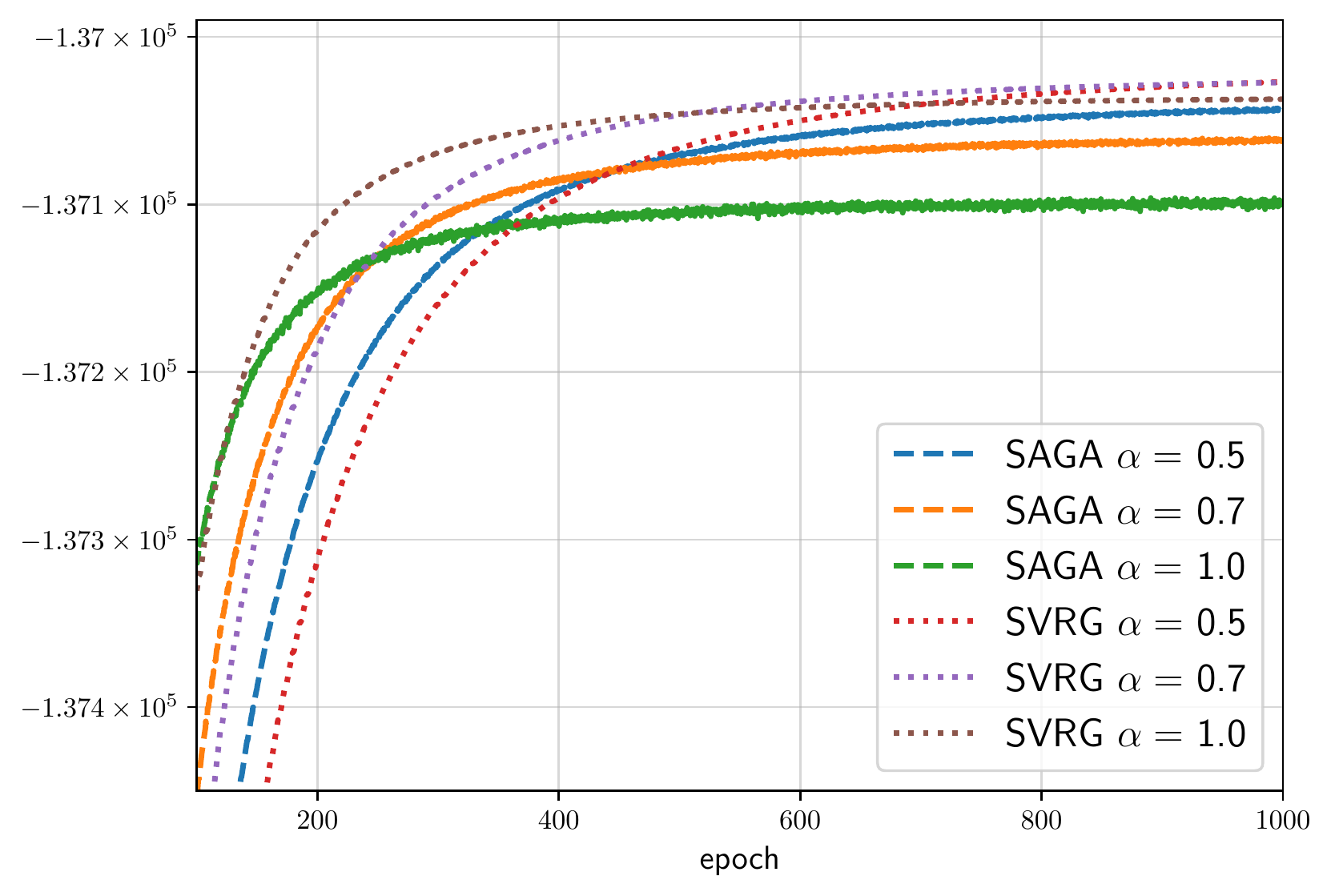} & \showpik{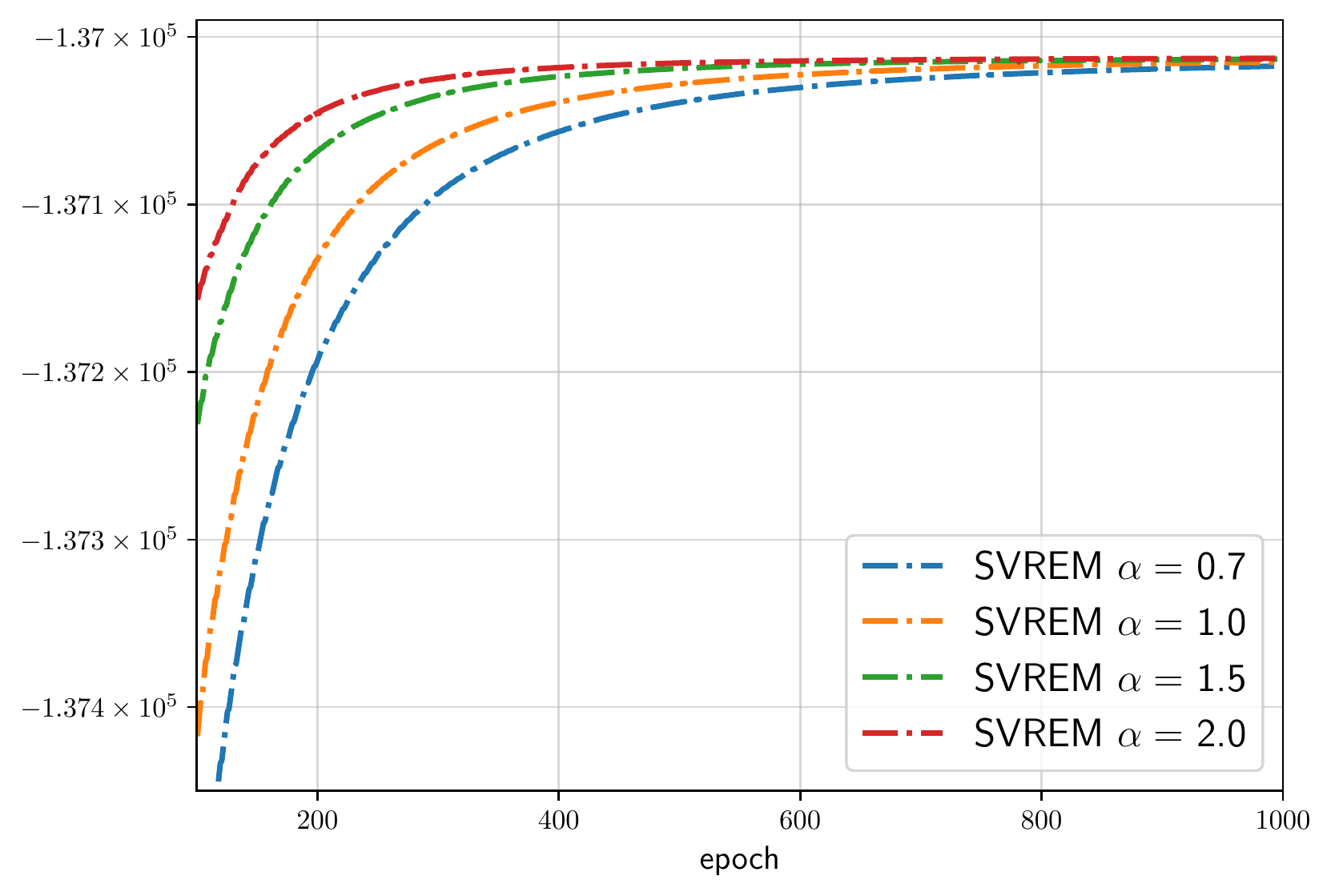} \\
(a) SAGA and SVRG & (b) SVREM
\end{tabular}
\caption{The influence of the stepsize $\alpha$ on SAGA, SVRG, and SVREM for the XCAT torso with 40 subsets and over $1000$ epochs. }\label{fig:Torso-longtime}
\end{figure}

\section{Conclusions}
In this work we revisited two classes of stochastic EM algorithms based on variance reductions from the machine learning community for the penalised PET reconstruction, one from the perspective of EM algorithms and the other from the perspective of preconditioned stochastic gradient ascent. Both classes of algorithms are straightforward to implement, and are applicable to a wide range of penalty terms.
The numerical results indicate that these algorithms  can effectively and efficiently accelerate the convergence of classical OSEM type algorithms, and hold significant potentials for the MAP problem in PET reconstructions.
This is particularly true for SVREM, which enjoys steady convergence towards the maximising solution, sometimes even with over-relaxation.
These promising empirical results naturally motivate further experimental evaluations on patient data and a theoretical analysis, to investigate the interplay between the number of subsets and the stepsize schedule (and also the "full-gradient" update frequency for SVRG), the role of over-relaxation in SVREM, and to precisely characterise their influence on the  convergence speed.
\appendix

\section{Proof of Theorem \ref{thm:sagasvrg_convergence}}\label{sec:proofs}

Note that the constrained maximisation problem
$$\argmax_{ f\geq 0} \Phi( f),\quad\mbox{with }\Phi( f):=\mathcal{L}( f) - \beta \mathcal{R}( f),$$
can be written as a minimisation problem
\[ \arg\min_{ f} -\Phi( f)+\chi_{\geq0}( f),\]
where $\chi_{\geq0}( f)$ is the indicator function of vectors with nonnegative entries.
It is proper, convex and lower semi-continuous and its proximal operator is the projection $P_{\geq0}$.
The conditions of the theorem imply
\begin{enumerate}
\item[(i)] Preconditioners $ d( f^{(k)})= d$ are constant, diagonal and positive.
\item[(ii)] $\Phi:\mathbb{R}^n\rightarrow\mathbb{R}$ is continuously differentiable
with a Lipschitz gradient, and $\Phi_t$ are continuously differentiable with an $L_t$-Lipschitz
gradient for each $t\in[N_s]$. Note that as is common with stochastic gradient algorithms, this may require to bound the projection operator to a finite box, whose bounds can be explicitly computed \cite{AF03} and which in practice affects neither the convergence nor the iterations, for which reason this is not presently included.
\item[(iii)] The set of maximisers is non-empty, i.e., $\argmax_{ f\geq0} \Phi( f)\neq\emptyset$.
\end{enumerate}
Then assumption (ii) holds for the modified objective $\widetilde\Phi( f):=\mathcal{L}( f) -
\beta \mathcal{R}( f)$, and the parameter $\varepsilon>0$ in \eqref{eqn:modified_LL} can be
chosen to be sufficiently small so that \cite{AF03}
$$\argmax_{ f\geq0} \Phi( f)=\argmax_{ f\geq0}
\widetilde\Phi( f)\quad\mbox{and}\quad \max_{ f\geq0} \Phi( f)=\max_{ f\geq0} \widetilde\Phi(f).$$
Below we use $\widetilde\Phi$ and identify $\widetilde\Phi=\Phi$. Motivated by \cite[Appendix C]{AF03}, we rewrite SAGA{} as
\begin{align}
\begin{split}
 \tilde f_{\rm saga}^{(k)} &=P_{\geq0}\bigg(\tilde f_{\rm saga}^{(k-1)} +\alpha\big(\widetilde \gdir^{(k)}_{t_k}- \widetilde \gdir^{(k-1)}_{t_k}+\frac{1}{N_s}\sum_{t=1}^{N_s}\widetilde \gdir^{(k-1)}_{t}\big)\bigg),
\end{split}
\end{align}
with
$$\tilde f^{(k)}=d^{-1/2} \odot f^{(k)},\quad \widetilde \gdir^{(k)}_{t}=\nabla\Psi_t(\tilde f^{(k)}),\mbox{ for }\Psi_t(\tilde f)=\Phi_t(d^{1/2}\odot\tilde f),$$
(with the entry-wise product). Now let $L=\max_{t\in[N_s]} L_t$.  Since $\Phi_t$ are $L$-Lipschitz and $d$ has non-negative entries, $\Psi_t$ are also Lipschitz, with a Lipschitz constant $L\|d^{1/2}\|_{\infty}=Ld_{\max}^{1/2}$.
Then with $\alpha=(3Ld_{\max}^{1/2})^{-1}$, by \cite[Theorem 2.1]{CLS18} (by changing maximisation to minimisation),
there exists an $\tilde f^\star\in\argmax_{ \tilde f\geq0}\Psi(\tilde f)$ such that $\Psi( \tilde f_{\rm saga}^{(k)})\rightarrow\Psi(\tilde f^\star)$ and
$\tilde f_{\rm saga}^{(k)}\rightarrow \tilde f^\star$ almost surely.

Similarly for SVRG, we have
\begin{align}
\begin{split}
 \tilde f_{\rm svrg}^{(k)} &=  P_{\geq0}\bigg(\tilde f_{\rm svrg}^{(k-1)} +\alpha\big(\nabla \Psi_{t_k}(\tilde f_{\rm svrg}^{(k-1)}) - \nabla \Psi_{t_k}(\tilde f^{\text{anc}})+\widetilde G\big)\bigg),
\end{split}
\end{align}
where $\widetilde G = \frac{1}{N_s}\nabla \Psi(\tilde f^{\text{anc}})$.
Then by \cite[Theorem 2.2.(i)]{CLS18}, with $\alpha\leq (4Ld_{\max}^{1/2}{(\ufrq N_s+2)})^{-1}$,
there exists an $\tilde f^\star\in\argmax_{\tilde f\geq0}\Psi(\tilde f)$ such that $\tilde f_{\rm svrg}^{(k)}\rightarrow \tilde f^\star$ almost surely.
Moreover, at the point when the full gradient is updated, i.e. if $k \mod \ufrq N_s = 0$, we have
$\mathbb{E}[\Psi(\tilde f^\star)-\Psi(\tilde f_{\rm svrg}^{(k)})]=\mathcal{O}(k^{-1})$.

By definition $\Psi( \tilde f)=\Phi( f)$, we have $\Phi( f_{\rm saga}^{(k)})\rightarrow\Phi( f^\star)$ and $ f_{\rm saga}^{(k)}\rightarrow f^\star$, and $ f_{\rm svrg}^{(k)}\rightarrow f^\star$ almost surely.
Moreover,  we have $\mathbb{E}[\Phi( f^\star)-\Phi( f_{\rm svrg}^{(k\ufrq N_s)})]=\mathcal{O}(1/k)$.

\section{The differences between SVREM{} and SVRG{} for ML problems}

It is tempting to think that SVRG{} and SVREM{} are equivalent in a certain context.
We compare the two algorithms for the ML (i.e., unpenalised) problem, which represents
the simplest case. Assume that both algorithms employ the same anchored estimate
$f^{\text{anc}}$, the full gradient, and the subset index $t$. For SVREM{}, we first
update the estimator $\widehat s^{(k)}$ of the full statistic by
\[
\widehat{\fss}^{(k+1)} = (1-\alpha)\widehat{\fss}^{k} +\alpha\big( \sss_{t_k}({\widehat f}_{\mathrm{svrem}}^{(k)}) - \sss_{t_k}({\widehat f}^{\text{anc}}) + t( f^{\text{anc}})\big),
\]
and then compute the maximising solution $f_{{\mathrm{svrem}}}^{(k)}$ by
\[
f_{{\mathrm{svrem}}}^{(k)} = \bigg(\frac{\widehat s^{(k)}}{\sum_{m=1}^Ma_{mn}}\bigg)_{n=1}^N =  f_{{\mathrm{svrem}}}^{(k)} + \alpha \sdir^{(k)},
\]
where the search direction $\sdir^{(k)}$ is given by
\begin{align*}
\frac{\sdir^{(k)}}{N_s}& =   d(f^{(k)})\odot\nabla\mathcal{L}_t( f^{(k)})-  d(f^{\text{anc}})\odot\nabla\mathcal{L}_t( f^{\text{anc}})+\frac{1}{N_s} d(f^{\text{anc}})\odot\nabla\mathcal{L}( f^{\text{anc}}) \\
 &\quad+ ( f^{(k)}- f^{\text{anc}})\odot \big(A_t^\top{1}\oslash A^\top{1} - \frac{1}{N_s}\big).
\end{align*}
Meanwhile, for SVRG{} updates with variable preconditioner, we have
\begin{align}
 f_{\rm svrg}^{(k+1)} =   f_{\rm svrg}^{(k)} +\alpha  d( f_{\rm svrg}^{(k)})\odot\bigg(\nabla \mathcal{L}_{t_k}( f_{\rm svrg}^{(k)}) - \nabla \mathcal{L}_{t_k}( f^{\text{anc}})+\frac{1}{N_s}\nabla \mathcal{L}( f^{\text{anc}})\bigg).
\end{align}
Thus, the two algorithms substantially differ even if SVRG uses a iteration 
dependent preconditioner $  d( f_{\rm svrg}^{(k)})= f_{\rm svrg}^{(k)}\oslash A^\top{1}$ (or indeed even a subset dependent preconditioner).
There are several notable differences. First, SVREM uses different preconditioners for each term, that is, the gradient terms containing $f^{\text{anc}}$ are preconditioned with $d(f^{\text{anc}})$, and $d(f^{(k)})$, for SVREM and SVRG, respectively. Second, the search direction for SVREM involves a term akin to momentum.

\bibliographystyle{abbrv}
\bibliography{svrem}
\end{document}